\documentclass[preprint]{elsarticle}

\usepackage{amsmath}
\usepackage{amsthm}
\usepackage{amsfonts}
\usepackage{derivative}
\usepackage{chngcntr}
\usepackage{graphicx}
\usepackage{caption}
\usepackage{subcaption}
\usepackage{mwe}
\usepackage{hyperref}
\usepackage{float}
\usepackage{geometry}
\usepackage{algorithm}
\usepackage{algpseudocode}
\usepackage[normalem]{ulem}

\makeatletter
\def\ps@first{%
   \let\@oddhead\@empty
   \let\@evenhead\@empty
   \def\@oddfoot{}
   \let\@evenfoot\@oddfoot
}

\newcommand{\Rea}{\mathbb{R}}

\newcommand{\paren}[1]{\left( #1 \right)}

\newcommand{\abs}[1]{\left| #1 \right|}
\newcommand{\norm}[1]{\left\lVert #1 \right\rVert}

\newcommand{\expect}[2]{\mathbb{E}_{#1}\left[#2\right]}
\newcommand{\braket}[2]{\left\langle #1 \middle| #2 \right\rangle}
\newcommand{\ket}[1]{\left| #1 \right\rangle}
\newcommand{\bra}[1]{\left\langle #1 \right|}

\DeclareMathOperator*{\argmin}{arg\,min}

\theoremstyle{definition}

\usepackage{mdframed}

\usepackage{xcolor}

\newcommand{\rev}[1]{#1}
\newcommand{\revB}[1]{#1}
\newcommand{\revC}[1]{#1}

\newcommand{\DeptMath}{Department of Mathematics, University of California, Berkeley, CA 94720, USA}
\newcommand{\Simons}{The Simons Institute for the Theory of Computing, Berkeley, CA 94720, USA}
\newcommand{\LBLMath}{Applied Mathematics and Computational Research Division, Lawrence Berkeley National Laboratory, Berkeley, CA 94720, USA}

\newcommand{\projsr}{SPRING}

\begin{document} 

\title{A Kaczmarz-inspired approach to accelerate the optimization of neural network wavefunctions}

\author[affi1]{Gil Goldshlager}
\author[affi2]{Nilin Abrahamsen}
\author[affi1,affi3]{Lin Lin}

\address[affi1]{\DeptMath}
\address[affi2]{\Simons}
\address[affi3]{\LBLMath}
\date{\today}

\begin{abstract}
Neural network wavefunctions optimized using the variational Monte Carlo method have been shown to produce highly accurate results for the electronic structure of atoms and small molecules, but the high cost of optimizing such wavefunctions prevents their application to larger systems. We propose the Subsampled Projected-Increment Natural Gradient Descent (SPRING) optimizer to reduce this bottleneck. SPRING combines ideas from the recently introduced minimum-step stochastic reconfiguration optimizer (MinSR) and the classical randomized Kaczmarz method for solving linear least-squares problems. We demonstrate that SPRING outperforms both MinSR and the popular Kronecker-Factored Approximate Curvature method (KFAC) across a number of small atoms and molecules, given that the learning rates of all methods are optimally tuned. For example, on the oxygen atom, SPRING attains chemical accuracy after forty thousand training iterations, whereas both MinSR and KFAC fail to do so even after one hundred thousand iterations.

\end{abstract}

\maketitle

\section{Introduction}
Predicting the properties of molecules and materials from first principles has numerous applications. For many chemical properties, it suffices to work within the {Born-Oppenheimer approximation}, in which the nuclei are viewed as classical point charges and only the electrons exhibit quantum-mechanical behavior. The study of chemistry through this lens is known as \emph{electronic structure theory}.

Within electronic structure theory, methods to model the many-body electron wavefunction include Hartree-Fock theory, configuration interaction methods, and coupled cluster theory.  A typical ansatz for such methods is a sum of Slater determinants which represent antisymmetrized products of single-particle states. The benefit of such an ansatz is that the energy and other properties of the wavefunction can be evaluated analytically from pre-computed few-particle integrals. 

Another approach to the electronic structure problem is the variational Monte Carlo method (VMC)  \cite{foulkes2001quantum, becca2017quantum}. In VMC, the properties of the wavefunction are calculated using Monte Carlo sampling rather than direct numerical integration, and the energy is variationally minimized through a stochastic optimization procedure. This increases the cost of the calculations, especially when high accuracy is required, but it enables the use of much more general ansatzes. Traditionally, these ansatzes included Slater-Jastrow wavefunctions and Slater-Jastrow-backflow wavefunctions \cite{foulkes2001quantum}.

In recent decades the modeling of very high-dimensional data and functions has seen impressive progress through the use of neural networks. For the quantum chemistry problem, the high-dimensional quantum wavefunction can be modeled by combining several neural network layers with a determinantal layer that enforces the Fermionic antisymmetry \cite{pfau2020ab,hermann2020deep}. Such neural network wavefunctions, optimized using the variational Monte Carlo method, have enabled near-exact computation of the ground-state energy for small molecules \cite{schatzle2023deepqmc, von2022self, gerard2022gold} and certain condensed matter systems \cite{Casella2023, kim2023neuralnetwork, lou2023neural, pescia2023messagepassing, li2022ab}.

Due to the large number of parameters and the highly nonlinear nature of neural networks, the optimization of neural network wavefunctions poses a significant challenge. Prior to neural network wavefunctions, VMC simulations typically relied on powerful domain-specific optimizers such as stochastic reconfiguration (SR) \cite{sorella_sr, nightingale2001optimization, sorella2007weak}  and the linear method \cite{ umrigar2007, toulouse2007optimization, neuscamman2012optimizing, zhao2017blocked, sabzevari2020accelerated}. These optimizers are able to converge in only tens or hundreds of iterations using highly accurate gradient estimates based on millions or more Monte Carlo samples. However, when applied to a wavefunction with $N_p$ parameters, they require solving either a linear system or a generalized eigenvalue problem involving a dense $N_p \times N_p$ matrix known as the $S$ matrix, with a cost that is $O(N_p^3)$ in general. \revC{The recently proposed Rayleigh-Gauss-Newton optimizer \cite{webber2022rayleigh} presents a potentially favorable intermediate point between stochastic reconfiguration and the linear method, but still incurs a similar computational cost.}

Since high-accuracy neural network architectures for molecules typically involve at least hundreds of thousands of parameters, it is not possible to directly apply SR or the linear method in this context. Instead, molecular neural network wavefunctions are typically optimized using methods from the machine learning community. These methods have a low per-iteration cost and require only a small number of Monte Carlo samples per iteration (or in machine learning parlance, a small minibatch size). However, they can require hundreds of thousands of optimization steps to converge fully.

For the FermiNet and related architectures \cite{pfau2020ab, von2022self, li2022ab, schatzle2023deepqmc}, the most popular optimizer is the Kronecker-Factored Approximate Curvature method (KFAC) \cite{martens2015optimizing}. KFAC was originally designed as a tractable approximation to natural gradient descent (NGD) \cite{amari1998natural} for machine learning models with millions of parameters. In the quantum setting, natural gradient descent is equivalent to stochastic reconfiguration since the $S$ matrix can be viewed as the Fisher information matrix of the normalized probability distribution defined by the wavefunction \cite{pfau2020ab}. Taking the view of stochastic reconfiguration, the key ingredient of KFAC is an approximate factorization of the $S$ matrix. This factorization enables the efficient inversion of the approximate $S$ matrix and also makes it possible to approximate $S$ based on multiple recent minibatches. This can significantly improve performance in highly stochastic settings where a single minibatch can provide only a very noisy approximation to the true $S$ matrix. It is worth noting that the interpretation of KFAC as an approximate natural gradient method has recently been called into question due to some experiments which show that KFAC performs \textit{better} than exact natural gradient updates \cite{benzing2022gradient}. Regardless of the underlying mechanism, KFAC represents the state-of-the-art in optimizing neural network wavefunctions for molecules and solids.

An alternative to KFAC is to take advantage of the fact that the estimate of the $S$ matrix is always low-rank when it is based on only a small set of Monte Carlo samples. This idea was first introduced in the machine learning community in the form of efficient subsampled natural gradient descent \cite{ren2019efficient}. In the physics community, several recent works have proposed similar methods  for the VMC setting \cite{minsr, rende2023simple}. These methods make fewer heuristic assumptions than KFAC and are much simpler to describe and implement. However, unlike KFAC, they are limited to estimating the $S$ matrix based on only a single minibatch at a time.

In this work, we improve upon these existing optimizers with a new method that we call Subsampled Projected-Increment Natural Gradient Descent (\projsr{}). Our method is most directly inspired by the minimum-step stochastic reconfiguration (MinSR) approach of Chen and Heyl \cite{minsr}. MinSR is based on the observation that the SR parameter update can be formulated as the solution to an overdetermined linear least-squares problem. At each iteration, the Monte Carlo samples provide access to a small set of rows from this least-squares problem, yielding an underdetermined subproblem which has many possible solutions. To make the parameter update unique, Chen and Heyl propose to choose the minimal-norm solution to the sampled subproblem. 

We improve upon this scheme by taking inspiration from the randomized Kaczmarz method for solving overdetermined linear least-squares problems \cite{strohmer2009randomized}. The original randomized Kaczmarz method solves the problem by sampling a single row at a time. The subsequently developed block version of the method (see i.e. \cite{needell2014paved}) instead uses a small set of rows at each iteration. In either case, the key idea of the Kaczmarz method is to iteratively project the solution vector onto the hyperplane of solutions to each sampled subproblem. When the system is consistent and the rows are sampled from an appropriate probability distribution, this approach provably converges to the true solution \revC{with an expected error that decays} by a constant factor at each iteration. 

Recall that the SR parameter update can be formulated as the solution to an overdetermined linear least-squares problem, and that each minibatch of Monte Carlo samples provides access to a randomly sampled underdetermined subproblem. In direct analogy to the Kaczmarz method, the parameter update in SPRING is calculated by projecting the \emph{previous} parameter update onto the hyperplane of solutions to the newly sampled subproblem. By leveraging the previous parameter update as the starting point for the projection, SPRING is able to use data from previous minibatches to obtain a more accurate approximation to the true SR update direction. Furthermore, this improvement is obtained at essentially no extra cost relative to MinSR.

To demonstrate the effectiveness of \projsr{}, we apply it to optimize a FermiNet wavefunction for several small atoms and molecules, namely the carbon, nitrogen, and oxygen atoms, two configurations of the N2 molecule, \revB{and the CO molecule at equilibrium}. We use the VMCNet code \cite{vmcnet2021github} for all numerical experiments. We find that \projsr{} consistently outperforms MinSR and KFAC when all three methods have their learning rates tuned for maximal performance. Remarkably, on the oxygen atom, \projsr{} is able to attain chemical accuracy after forty thousand training iterations, whereas both MinSR and KFAC fail to do so even after one hundred thousand iterations.

\section{Background}

\subsection{Variational Monte Carlo}
\newcommand{\nuc}[1]{\revB{\mathsf{R}_{#1}}}

In this work we focus on the application of the variational Monte Carlo method to find ground-state wavefunctions for molecular systems. A molecule is defined by the positions $\nuc{I}$ and charges $Z_I$ of $M$ atomic nuclei. Using the Born-Oppenheimer approximation \cite{born1927}, we model the nuclei as fixed point charges. The quantum system of interest then consists of $N$ electrons interacting under the influence of these charges.  The Hamiltonian for the system is given in atomic units by
\begin{equation}
H = - \frac{1}{2} \revB{\sum_{i=1}^N \Delta_{r_i}} - \sum_{i = 1}^N \sum_{I=1}^M \frac{Z_I}{\abs{r_i - \nuc{I}}} + \sum_{i<j}^N \frac{1}{\abs{r_i - r_j}} + \sum_{I < J}^M \frac{Z_I Z_J}{\abs{\nuc{I} - \nuc{J} }},
\end{equation}
\revB{where $r_i \in \Rea^3$ represents the position of electron $i$ and $\Delta_{r_i}$ represents the corresponding Laplacian operator}. Because electrons are fermions, the wavefunction must be antisymmetric with respect to particle exchange. Within the space of antisymmetric wavefunctions, the ground-state energy $E_0$ and wavefunction $\ket{\psi_0}$ correspond to the smallest eigenvalue of $H$ and its corresponding eigenfunction.

We make several simplifications to the space of wavefunctions before searching for the ground-state. First, since the Hamiltonian is Hermitian, its ground-state wavefunction can be chosen to be purely real. We thus limit our search to real normalizable many-body wavefunctions of the form $\psi:\Rea^{3N} \rightarrow \Rea$, usually written as $\psi(R) = \psi(r_1, \ldots, r_N)$ with $R \in \Rea^{3N}$. Next, we fix the numbers $N_\uparrow, N_\downarrow$ of up- and down-spin electrons \textit{a priori} for each calculation. Given this assumption, together with the antisymmetry of the wavefunction and the spin-independence of the Hamiltonian, it is possible to assume without loss of generality that the first $N_\uparrow$ electrons are always spin-up and the last $N_\downarrow$ electrons are always spin-down \cite{foulkes2001quantum}.

The ground-state wavefunction can be found by minimizing the expectation value of the energy:
\begin{equation}
\ket{\psi_0} = \underset{\psi}{\operatorname{argmin}} \frac{\bra{\psi}H\ket{\psi}}{\braket{\psi}{\psi}}.
\end{equation}
Given a variational ansatz $\ket{\psi_\theta}$, we can then define a loss function 
\begin{equation}
L(\theta) = \frac{\bra{\psi_\theta}H\ket{\psi_\theta}}{\braket{\psi_\theta}{\psi_\theta}}
\end{equation}
and represent the ground-state approximately via
\begin{equation}
\theta^* = \underset{\theta}{\operatorname{argmin}}\; L(\theta),\; \ket{\psi_0} \approx \ket{\psi_{\theta^*}},\; E_0 \approx L(\theta^*).
\end{equation}

In some electronic structure methods, such as the Hartree-Fock method and configuration interaction methods, the ansatz is chosen to enable the direct calculation of $L(\theta)$ based on precomputed one- and two-electron integrals. In variational Monte Carlo, the ansatz is chosen in a more flexible way, such as a Slater-Jastrow ansatz, a Slater-Jastrow-backflow ansatz, or a neural network wavefunction. For such ansatzes it is not possible to directly evaluate $L(\theta)$, so it is necessary to instead approximate it via Monte Carlo integration. To this end, $L(\theta)$ can be reformulated stochastically as
\begin{equation}
L(\theta) = \expect{R \sim p}{E_L(\theta)},
\end{equation}
where $p(R) = \psi_\theta(R)^2 / \braket{\psi_\theta}{\psi_\theta}$ and $E_L(R) = H\psi_\theta(R) / \psi_\theta(R)$ (see i.e. \cite{foulkes2001quantum}, Section III.C). This last term, $E_L(R)$, is known as the \textit{local energy} of the wavefunction $\ket{\psi_\theta}$ at the position $R$. 

To optimize a variational wavefunction using this formula, it is usually necessary to calculate the gradient $\nabla_\theta L(\theta)$. A convenient formula for this gradient can be derived taking advantage of the fact that $H$ is Hermitian (see i.e. \cite{lin2023explicitly}, Appendix E):
\begin{equation}
g = \nabla_\theta L(\theta) = 2\expect{R \sim p}{\nabla_\theta \log \abs{\psi_\theta(R)} \paren{E_L(R) - L(\theta)}}.
\end{equation}
In practice, both $L(\theta)$ and $g$ are estimated stochastically  using Markov-Chain Monte Carlo sampling to generate samples from $p(R)$. The parameters can then be updated by a variety of optimization schemes, several of which we discuss in the following sections.

\subsection{Stochastic Reconfiguration}
One popular approach for optimizing variational wavefunctions is known as stochastic reconfiguration (SR) \cite{sorella_sr, nightingale2001optimization, sorella2007weak} . In recent years, SR has been widely adopted for optimizing neural quantum states in second quantization \cite{carleo2017solving, carleo2019netket}. It also serves as the starting point for the MinSR method to be introduced later.


SR is based on the idea of imaginary time evolution, which states that the ground-state $\ket{\psi_0}$ can be found via the formula
\begin{equation}
\ket{\psi_0} = \lim_{\tau \rightarrow \infty} e^{-\tau H} \ket{\psi}
\end{equation}
for any wavefunction $\ket{\psi}$ such that $\braket{\psi_0}{\psi} \neq 0$. In SR, we optimize a variational ansatz $\ket{\psi_\theta}$ by finding parameter updates that approximate small imaginary time steps. In particular, we choose a step-size $\delta \tau$ and use this to define the desired updated wavefunction $\ket{\psi'} = e^{-\delta \tau H} \ket{\psi_\theta}$. Due to the constraints of the ansatz, we cannot update our wavefunction to $\ket{\psi'}$ directly. Instead we seek a parameter update $d\theta$ such that $\ket{\psi_{\theta + d\theta}}$ is close to $\ket{\psi'}$. To measure the closeness, we rely on the Fubini-Study distance defined by
\begin{equation}
D(\ket{\psi}, \ket{\phi}) = \operatorname{arccos} \frac{\abs{\braket{\psi}{\phi}}}{\norm{\psi} \norm{\phi}}.
\end{equation}
In practice, we make a first order approximation to both $\ket{\psi'}$ and $\ket{\psi_{\theta + d\theta}}$ and then minimize a second-order approximation to $D^2(\ket{\psi'}, \ket{\psi_{\theta + d\theta}})$. See \cite{park2020geometry} for a more thorough discussion of the geometric underpinnings of stochastic reconfiguration.

To present the formula for the update, we first define the functions $\bar{O}(R)$ and $\bar{\epsilon}(R)$. The first quantity, $\bar{O}(R)$, is the transposed gradient  of the logarithm of the normalized wavefunction, evaluated at the point $R$. The second quantity, $\bar{\epsilon}(R)$, is the first-order change in the logarithm of the normalized wavefunction induced by the imaginary time-step $\delta \tau$, evaluated at the point $R$. These quantities are given by 
\begin{equation}
\bar{O}(R) = \paren{\nabla_\theta \log \abs{\psi_\theta(R)} - \expect{R \sim p}{ \nabla_\theta \log \abs{\psi_\theta(R)}}}^T,
\end{equation} 
\begin{equation}
\bar{\epsilon}(R) = \revC{-\delta \tau \paren{E_L(R) - \expect{R \sim p}{E_L(R)}}}.
\end{equation}
It can be shown that, to second order, the square of the Fubini-study distance is given by
\begin{equation}
D^2(\ket{\psi'}, \ket{\psi_{\theta + d\theta}}) = \expect{R \sim p}{\abs{\bar{O}(R)d\theta - \bar{\epsilon}(R)}^2}. \label{eq:leastsquares}
\end{equation}
This leads us to define the SR parameter update by
\begin{equation}
d\theta = \underset{d\theta'}{\operatorname{argmin}}\; \expect{R \sim p}{\abs{\bar{O}(R)d\theta' - \bar{\epsilon}(R)}^2}. \label{eq:sr_system}
\end{equation}
A detailed derivation of this formulation of stochastic reconfiguration can be found in \cite{minsr}.

In practice, (\ref{eq:leastsquares}) must always be approximated using a finite collection of $N_s$ samples $R_1, \ldots, R_{N_s}$. \revC{To estimate $\bar{O}(R)$ and $\bar{\epsilon}(R)$ at the sampled points, we must empirically estimate the expectation values that appear in their definitions. We first define \begin{equation}
O = \frac{1}{\sqrt{N_s}} \begin{bmatrix} \nabla_\theta \log \abs{\psi_\theta(R_1)} \\
\vdots \\ 
 \nabla_\theta \log \abs{\psi_\theta(R_{N_S})} \end{bmatrix},\; \epsilon = - \frac{\delta \tau}{\sqrt{N_s}}\begin{bmatrix}
E_L(R_1) \\ \vdots \\ E_L(R_{N_s})
\end{bmatrix}.
\label{eq:Oeps}
\end{equation}
Next, let $\mathbf{1}$ be the column vector of length $N_S$ whose entries are all equal to $1$ and let $P = \frac{1}{N_s} \mathbf{1} \mathbf{1}^T$ represent the orthogonal projector onto the span of $\mathbf{1}$. When acting on the left, the operator $P$ replaces each row of the matrix or vector on its right with the average of all of its rows. Our empirical estimates of $\bar{O}(R)$ and $\bar{\epsilon}(R)$ can thus be collected as
\begin{equation}
\bar{O} = (I-P) O,\; \bar{\epsilon} = (I-P) \epsilon
\label{eq:Oeps_bar}.
\end{equation}} The estimate of the Fubini-Study distance of the update $d\theta$ can then be written as
\begin{equation}
D^2(\ket{\psi'}, \ket{\psi_{\theta + d\theta}}) \approx \norm{\bar{O} d\theta - \bar{\epsilon}}^2,
\end{equation}
and the update takes the form
\begin{equation}
d\theta = \underset{d\theta'}{\operatorname{argmin}}\; \norm{\bar{O} d\theta' - \bar{\epsilon}}^2 \label{eq:sr_least}.
\end{equation}

In traditional SR, many more samples are taken than parameters in the ansatz, meaning that the least-squares problem is overdetermined and should have a unique solution. To protect against the case when $\bar{O}$ is ill-conditioned or singular, a Tikhonov regularization is generally added to yield the regularized problem
\begin{equation}
d\theta = \underset{d\theta'}{\operatorname{argmin}}\; \frac{1}{\lambda} \norm{\bar{O} d\theta' - \bar{\epsilon}}^2 + \norm{d\theta'}^2 \label{eq:sr_reg}.
\end{equation}
In the case of real parameters and a real wavefunction, the solution to (\ref{eq:sr_reg}) is given by
\begin{equation}
d\theta = (\bar{O}^T \bar{O} + \lambda I)^{-1} \bar{O}^T \bar{\epsilon}. \label{eq:sr_sol}
\end{equation}
This is more traditionally written as 
\begin{equation}
d\theta = \revC{-}\frac{\delta \tau}{2} (S + \lambda I)^{-1}g, 
\end{equation}
where $S = \bar{O}^T \bar{O}$, $g = \nabla_\theta L(\theta) = \revC{-} 2\bar{O}^T \bar{\epsilon}/ \delta \tau$, and the imaginary time step has been halved to cancel the factor of two in the gradient.

\subsection{Minimum-Step Stochastic Reconfiguration \label{sec:back:minsr}}
For a problem with $N_p$ variational parameters, the traditional SR update scales as $O(N_p^3)$ due to the need to invert the $N_p \times N_p$ matrix $S$.  The insight of MinSR is that the parameter update can be calculated much more efficiently when $N_s \ll N_p$. In fact the linear system is highly underdetermined in this setting. The MinSR method addresses this issue by choosing the solution with minimal norm. For regularization, MinSR utilizes a pseudoinverse with a cutoff for small eigenvalues. Subsequently, Rende et al \cite{rende2023simple} suggested using a Tikhonov regularization instead. This approach amounts to solving the same regularized equation
as in traditional SR, namely (\ref{eq:sr_reg}). In the underdetermined setting, the application of the Sherman-Morrison-Woodbury formula to (\ref{eq:sr_sol}) yields the solution
\begin{equation}
d\theta = \bar{O}^T (\bar{O} \bar{O}^T + \lambda I)^{-1} \bar{\epsilon}.
\end{equation}
This is the version of MinSR that we use as the basis for SPRING. The key point is that the update can now be calculated efficiently since $\bar{O} \bar{O}^T$ is only $N_s \times N_s$ rather than $N_p \times N_p$. 

It is worth noting that methods very similar to MinSR were introduced independently in the machine learning community as early as 2019. For example, Zhang, Martens, and Grosse decribed a pseudoinverse-based approach to natural gradient descent for the overparameterized case when the number of parameters is greater than the total number of data-points available \cite{zhang2019fast} . Subsequently, Ren and Goldfarb proposed an efficient subsampled natural gradient method  \cite{ren2019efficient} for the case when the minibatch size is smaller than the number of parameters. In fact, the Tikhonov-regularized form of MinSR can be viewed as a streamlined implementation of the method of Ren and Goldfarb which applies whenever the gradient of the loss function is a linear combination of the model gradients at the sampled points. We elaborate on this connection in \ref{app:NGD-SMW}.

\subsection{\rev{Kaczmarz method for solving linear systems}}
\rev{MinSR enables the efficient solution of an underdetermined subsample from the SR equation, but the solution of such an equation is not necessarily a good approximation to the solution of the original SR equation. Our goal is to develop an  algorithm that can converge to the solution of the original SR equation using only a series of underdetermined subsamples. This is the motivation behind the Kaczmarz algorithm, and in particular its blocked variants.} 

\rev{To understand the Kaczmarz algorithm, consider an overdetermined system of linear equations $Ax=b$, where $A$ is $m \times n$ and
\begin{equation}
A = \begin{bmatrix}
a_1^T \\
\vdots \\
a_m^T
\end{bmatrix},\; b = \begin{bmatrix}
b_1 \\
\vdots \\
b_m
\end{bmatrix}.
\end{equation}
Assume that $A$ has full column rank and that the system is consistent, meaning that there exists a unique $x^*$ such that $Ax^* = b$. The earliest form of the Kaczmarz algorithm, which dates back to the first half of the twentieth century \cite{karczmarz1937angenaherte}, solves this problem iteratively. A starting guess $x_0$ is required for initialization and is then honed by iterations of the form 
\begin{equation}
x_k = x_{k-1} + \frac{a_i}{\norm{a_i}^2}(b_i - a_i^T x_{k-1}),
\end{equation}
where $i$ is chosen to cycle through all the rows of $A$ one at a time. The interpretation is that $x_k$ is attained by projecting $x_{k-1}$ onto the solution space of the sampled equation $a_i^T x = b_i$.
}

\rev{
The seminal work of Strohmer and Vershynin \cite{strohmer2009randomized} proposed a randomized variant of the algorithm which samples row $i$ with a probability proportional to $\norm{a_i}^2$. With this randomized sampling strategy, they showed that the algorithm converges as
\begin{equation}
\expect{}{\norm{x_k - x^*}^2} \leq \paren{1 - \kappa_{D}(A)^{-2}}^k \norm{x_0 - x^*}^2,
\end{equation}
where $\kappa_{D}(A) = \norm{A}_F \norm{A^+}_2$ is the Demmel condition number of $A$ and  $A^+$ is the Moore-Penrose pseudoinverse of $A$. 
}

\newcommand{\samp}{\sigma}

\rev{
Since single row updates are very inefficient on modern hardware, blocked versions of the randomized Kaczmarz algorithm were subsequently proposed and analyzed, i.e. \cite{needell2014paved}. Such randomized block Kaczmarz algorithms take the form
\begin{equation}
x_k = x_{k-1} + A_\samp^+ (b_\samp - A_\samp x_{k-1}),
\end{equation}
where $\samp$ represents a randomly selected subset of the rows of $A$ and $b$ and $A_\samp^+$ represents the pseudoinverse of $A_\samp$. The interpretation is similar to before: $x_k$ represents the projection of $x_{k-1}$ onto the solution space of the sampled equation $A_\samp x = b_\samp$. For consistent equations the randomized block Kaczmarz algorithm converges linearly to the true solution $x^*$, though the convergence rate is more complicated than in the single row case and has a more nuanced dependence on the sampling procedure. 
}

\section{Methods \label{sec:meth}}

We now present our main contribution, the Subsampled Projected-Increment Natural Gradient Descent (SPRING) algorithm for optimizing neural network wavefunctions. To start, recall the formula for the SR parameter update: 
\begin{equation} \label{eq:SR_meth}
d\theta = \underset{d\theta'}{\operatorname{argmin}}\; \expect{R \sim p}{\abs{\bar{O}(R)d\theta' - \bar{\epsilon}(R)}^2}.
\end{equation}
Since $R$ is drawn from a continuous and high-dimensional space this equation can be seen as a highly overdetermined linear least-squares problem, with each configuration $R$ corresponding to a single row. \revC{Now, let $\bar{O}_k$ and $\bar{\epsilon}_k$ denote the subsamples from $\bar{O}(R)$ and $\bar{\epsilon}(R)$ that are available at VMC training iteration $k$. Our approach hinges on the} assumption that the imaginary time-step is small, meaning the parameter vector $\theta$ changes only slightly at each iteration. Thus, although the sampled subproblems  $(\bar{O}_{k-1}, \bar{\epsilon}_{k-1})$, $(\bar{O}_k, \bar{\epsilon}_k)$ can be completely different, the underlying \revC{SR equations are nearly identical.} This \revC{inspires us to proceed as follows:}
\begin{enumerate}
\item Make the approximation that the recent iterates $(\bar{O}_k,\bar{\epsilon}_k), (\bar{O}_{k-1},\bar{\epsilon}_{k-1}), \ldots$ are all random block samples from the current SR equation 
\revB{
\begin{equation}
\underset{d\theta'}{\operatorname{argmin}}\; \expect{R \sim p}{\abs{\bar{O}(R)d\theta' - \bar{\epsilon}(R)}^2}.
\end{equation}}
\item Apply the randomized block Kazcmarz method to these iterates to yield an \revB{approximate solution $\phi_k$ to the current SR equation.}
\revB{\item Apply some scaling and clipping procedures, to be described later, to $\phi_k$ to calculate the parameter update $d\theta_k$.}
\end{enumerate}
In this way SPRING takes advantage of the entire optimization history to inform each parameter update, distinguishing it from MinSR which leverages only a single minibatch of data at a time. 

\rev{Following the structure of a single step of the randomized block Kaczmarz method \cite{needell2014paved}, we calculate $\phi_k$ by projecting $\phi_{k-1}$ onto the solution space of the underdetermined equation $\bar{O}_k \phi = \bar{\epsilon}_k$. Recall that the basic form for the Kaczmarz projection is
\begin{equation}
\phi_k = \phi_{k-1} + \bar{O}_k^+ (\bar{\epsilon}_k - \bar{O}_k \phi_{k-1}).
\end{equation}
Equivalently we can say that $\phi_k$ satisfies 
\begin{equation}
\phi_k = \argmin_{\phi} \norm{\phi - \phi_{k-1}}^2 \text{ s.t. } \bar{O}_k \phi = \bar{\epsilon}_k.
\end{equation}
Thus, at a conceptual level, our algorithm} differs from MinSR only in how we break the indeterminacy of the sampled subproblem: MinSR chooses the solution of minimal norm, whereas SPRING chooses the solution that is nearest to the previous \rev{approximate solution}. 

\rev{In practice $\bar{O}_k$ can be singular or very ill-conditioned, so it is beneficial add some form of regularization to this projection. We choose to regularize the projection by incorporating the linear equations using a penalty term rather than a hard constraint. To make this modification, we introduce a regularization parameter $\lambda$ and use the formula 
\begin{align}
\phi_k = \underset{\phi}{\operatorname{argmin}}\; \frac{1}{\lambda} \norm{\bar{O}_k \phi - \bar{\epsilon}_k}^2 + \norm{\phi -\phi_{k-1}}^2. \label{eq:\projsr{}_nomu}
\end{align}}

\rev{Even with this regularization}, we find that directly using (\ref{eq:\projsr{}_nomu}) \revC{can result in unstable optimization trajectories (see Figure \ref{fig:mu})}. We do not yet have an explanation for this phenomenon. To stabilize the method, we decay the previous gradient by a small amount before projecting it. Formally, this stabilization comes in the form of a new regularization parameter $\mu$ and a modified update formula
\begin{align}
\revB{\phi_k} = \underset{\revB{\phi}}{\operatorname{argmin}}\; \frac{1}{\lambda} \norm{\bar{O}_k \revB{\phi} - \bar{\epsilon}_k}^2 + \norm{\revB{\phi} - \mu \revB{\phi}_{k-1}}^2. \label{eq:projsr}
\end{align}

An explicit formula for \revB{$\phi_k$} can now be derived. We first define \revB{$\pi_k = \phi_k - \mu \phi_{k-1}$, $\pi = \phi - \mu \phi_{k-1}$, and $\bar{\zeta}_k = \bar{\epsilon}_k - \mu \bar{O}_k \phi_{k-1}$} and note that (\ref{eq:projsr}) can be recast as 
\begin{align}
\revB{\pi_k} = \underset{\revB{\pi}}{\operatorname{argmin}}\; \frac{1}{\lambda} \norm{\bar{O}_k \revB{\pi} - \revB{\bar{\zeta}_k}}^2 + \norm{\revB{\pi}}^2.
\end{align}
This formula now has the same form as MinSR with Tikhonov regularization, so we know that the solution is given by 
\begin{equation}
\revB{\pi_k} =  \bar{O}_k^T (\bar{O}_k \bar{O}_k^T + \lambda I)^{-1} \rev{\bar{\zeta}_k}.
\end{equation}
Finally, we can translate back to \revB{$\phi_k$} yielding
\begin{equation}
\revB{\phi_k}  =\bar{O}_k^T (\bar{O}_k \bar{O}_k^T + \lambda I)^{-1} \rev{\bar{\zeta}_k} + \mu \revB{\phi_{k-1}}. \label{eq:spring}
\end{equation}
\rev{
To understand this formula, it is helpful to explicitly write out the three terms: 
\begin{equation}
\rev{\phi_k} = \bar{O}_k^T (\bar{O}_k \bar{O}_k^T + \lambda I)^{-1} \bar{\epsilon}_k + \mu \phi_{k-1}  - \mu \bar{O}_k^T (\bar{O}_k \bar{O}_k^T + \lambda I)^{-1} \bar{O}_k \phi_{k-1}. 
\end{equation}
The first term alone corresponds to MinSR, while the first two terms together correspond to a scaled version of MinSR with momentum (to be presented in Section \ref{sec:minsr+m}). It is the addition of the third term that concretely distinguishes SPRING from these other methods. 
}

\revB{Equation (\ref{eq:spring})} is almost the formula that we use in practice. The only missing \revB{piece is an extra stabilization procedure that we introduce in Section \ref{sec:T}}, which is not essential but should be included for best performance. \revB{Finally, to calculate $d\theta_k$}, we also include a learning rate schedule $\eta_k$ \revB{and a norm constraint $C$ which ensures that the parameters are not changed too much at each iteration (see Section \ref{sec:norm})}. \revB{To ensure that the Kaczmarz algorithm is implemented consistently, regardless of the learning rate schedule, we always calculate $\bar{\epsilon}_k$ using $\delta \tau=1$ and then apply the learning rate $\eta_k$ only when we calculate $d\theta_k$.} For convenience we present the full procedure for SPRING including these extra details in Algorithm \ref{alg:SPRING}. 

\begin{algorithm}
\caption{SPRING} \label{alg:SPRING}
\begin{algorithmic}[1]
\Require Hamiltonian $H$, ansatz $\psi_\theta$, 
\Require Initialization $\theta_0$, iteration count $K$, batch size $N_s$, learning rate schedule $\eta_k$
\Require Tikhonov damping $\lambda$, decay factor $\mu$, norm constraint $C$
\State $\theta \gets \theta_0$
\State $\revB{}{\phi} \gets 0$
\For{k in \revB{1:K}}
    \State Sample $R_1, \ldots, R_{N_s}$ from $p(R) = \psi_\theta(R)^2 / \braket{\psi_\theta}{\psi_\theta}$ \Comment{i.e. using MCMC}
    \State Calculate $\bar{O}$, $\bar{\epsilon}$ using $\delta \tau = 1$  \Comment{\revC{via (\ref{eq:Oeps}), (\ref{eq:Oeps_bar})}}
    \State $\bar{\zeta} \gets \bar{\epsilon} - \mu \bar{O} \phi$
    \State  $\revB{\phi} \gets \bar{O}^T (\bar{O}\bar{O}^T + \lambda I + \revB{\frac{1}{N_s} \mathbf{1} \mathbf{1}^T})^{-1}\revB{\bar{\zeta}}  + \mu \revB{\phi}$ \Comment{following (\ref{eq:TP}) and using Cholesky}
    \State $\revB{d\theta \gets \phi \cdot  \min(\eta_k, \sqrt{C} / \norm{\phi})}$ \Comment{See Section \ref{sec:norm}}
    \State $\theta \gets \theta + \revB{d\theta}$
\EndFor \\
\Return $\theta$
\end{algorithmic}
\end{algorithm}

It is worth noting that due to several technical details, we cannot directly transfer the convergence results for the classical Kaczmarz method to the VMC setting. First of all, the inclusion of the parameter $\mu$ in SPRING distinguishes it substantially from the Kaczmarz method. We can understand this distinction as arising from the fact that in \projsr{}, the underlying least-squares system changes a small amount after each iteration. Thus, it makes sense to progressively forget the information from previous iterations, which is what we accomplish by setting $\mu < 1$. Additionally, the least-squares problem arising in VMC need not be consistent, and in the inconsistent setting the Kaczmarz method only converges to within a ball of the true solution, with the radius of the ball depending on how inconsistent the system is \cite{needell2010randomized}. For these reasons, we cannot derive a rigorous convergence guarantee for SPRING with regard to either $d\theta$ or $\theta$. Still, the connection to the Kaczmarz method serves to motivate the algorithm and explain its superior performance relative to MinSR.

\subsection{Norm Constraint \label{sec:norm}}

As an additional stabilization procedure, we include a constraint on the norm of the MinSR, MinSR+M, and \projsr{} updates inspired by the norm constraint used by KFAC. We have found that including this norm constraint helps to stabilize the optimization procedure and enable larger learning rates to be applied, and it also improves the ability to transfer learning rates from one system to another. For reference we briefly describe how the norm constraint used by KFAC works before introducing the form of the norm constraint that we use for MinSR and \projsr{}.

The norm constraint in KFAC is ``natural'' in the sense that it applies to the norm induced by the Fisher information matrix as opposed to the Euclidean norm. To define this scheme let $g$ be the estimated energy gradient, $\eta$ be the learning rate, $\revC{S}$ be KFAC's approximation to the Fisher information matrix, and $\revB{\phi} = \revC{S}^{-1}g$ be the \revB{unscaled and unconstrained} KFAC parameter update. \revB{The final update will be $d\theta = c \phi$ for some scalar $c$.} The norm constraint relies upon the fact that \revB{$\norm{d\theta}_{\revC{S}}^2 = c^2 \phi^T \revC{S} \phi = c^2 \phi^T g$} and as a result it enforces the constraint
\begin{equation}
c^2 \phi^T g \leq C.
\end{equation}
\revB{
Concretely this is achieved by choosing 
\begin{equation}
d\theta = \phi \cdot \min(\eta, \sqrt{C} / \sqrt{\phi^T g}).
\end{equation}
}

In \projsr{} the update $d\theta$ and the gradient $g$ are not related by a positive definite matrix \revC{S}; hence the above scheme is not justified. Instead we implement a simple Euclidean norm constraint given by
\begin{equation}
\norm{d\theta}^2 \leq C.
\end{equation}
\revB{
Just like in KFAC, if this constraint is violated before the norm constraint is applied, we scale the update down to ensure that it is satisfied. Concretely, we calculate the update $d\theta_k$ from the Kaczmarz iterate $\phi_k$ and the learning rate $\eta_k$ using the formula 
$$d\theta_k = \phi_k \cdot  \min(\eta_k, \sqrt{C} / \norm{\phi_k}).$$}

\subsection{Numerically Stable Inversion \label{sec:T}}
In exact arithmetic, $T = \bar{O} \bar{O}^T$ is positive semidefinite, $T + \lambda I$ is positive definite, and $(T + \lambda I)^{-1} \rev{\bar{\zeta}}$ can be computed efficiently using the Cholesky decomposition. However, in single precision we have found that the $T$ matrix can be become indefinite. As a result, the Cholesky decomposition can fail if the regularization is chosen to be too small, and we have even observed this to occur when using our default value of $\lambda = 0.001$.

We now note that $T$ is \textit{always} singular due to the construction of $\bar{O}$. \revC{Recall that $\mathbf{1}$ represents} the column vector whose entries are all equal to $1$ and $P = \frac{1}{N_s} \mathbf{1} \mathbf{1}^T$ represents the orthogonal projector onto the span of $\mathbf{1}$. \revC{Since} $\bar{O} = \paren{I - P} O$, we have that
\begin{equation}
T \mathbf{1} = \bar{O} \bar{O}^T \mathbf{1} = \bar{O} O^T \paren{I - P} \mathbf{1} =  \bar{O} O^T  (\mathbf{1} - \mathbf{1}) = 0.
\end{equation}
This means that any numerical perturbation to $T$ can make it indefinite. If the numerical perturbation is large enough, the resulting $T$ matrix can have an eigenvalue smaller than $-\lambda$, in which case the Cholesky factorization fails. This can happen even when all of the other eigenvalues of $T$ are well-separated from $0$.

We alleviate this issue by replacing the matrix $T + \lambda I$ with the further regularized matrix \rev{$T + \lambda I + \omega P$ for some positive real $\omega$}. We thus calculate the \rev{Kaczmarz iterate} for iteration $k$ as
\rev{
\begin{equation}
\rev{\phi_k}  =\bar{O}_k^T (T_k + \lambda I + \omega P)^{-1} \rev{\bar{\zeta}_k} + \mu \phi_{k-1}.
\label{eq:TP}
\end{equation}
}
This does not affect the value of \rev{$\phi_k$} in exact arithmetic. To see this, note that $\mathbf{1}$ is an eigenvector of $(T_k + \lambda I)$ with eigenvalue $\lambda$, and is thus also an eigenvector of \rev{$(T_k + \lambda I + \omega P)$ with eigenvalue $\omega + \lambda$}. As a result, we have
\rev{
\begin{equation}
(T_k + \lambda I + \omega P)^{-1} = (T_k + \lambda I)^{-1} + \paren{\frac{1}{\omega+\lambda} - \frac{1}{\lambda}} P.
\end{equation}
}
Acting on the left with $\bar{O}_k^T$ then annihilates the extra term: 
\rev{\begin{align*}
\bar{O}_K^T (T_k + \lambda I + \omega P)^{-1} &= \bar{O}_k^T (T_k + \lambda I)^{-1} + \paren{\frac{1}{\omega+\lambda} - \frac{1}{\lambda}}  \bar{O}_k^T P \\
&= \bar{O}_k^T (T_k + \lambda I)^{-1}.
\end{align*}}

Thus, using (\ref{eq:TP}) reduces the likelihood that numerical errors cause the Cholesky solver to fail without otherwise affecting the computation. \rev{In our experiments we have used the value $\omega=1$ in all cases. However, it is theoretically possible for the eigenvalues of $T$ to all be smaller than $1$, in which case setting $\omega=1$ would increase the condition number of the regularized $T$ matrix. In such a case $\omega$ could instead be chosen for example to be the mean of the  eigenvalues of the original $T$ matrix.}

\subsection{\revB{MinSR implementation details}}
\revB{
We now briefly discuss the version of MinSR that we use for our numerical experiments, which is very similar but not identical to the basic version presented in Section \ref{sec:back:minsr}. Functionally, the only differences are the inclusion of a norm constraint and the extra regularization of  the matrix $\bar{O}_k \bar{O}_k^T$ as described in Section \ref{sec:T}. Additionally, for consistency with SPRING we set $\delta \tau = 1$ when calculating $\bar{\epsilon}_k$ and we instead control the step size with a learning rate parameter $\eta_k$.  With these modifications the MinSR algorithm is described by the following formulas:
\begin{equation}
\phi_k = \bar{O}_k^T (\bar{O}_k \bar{O}_k^T + \lambda I + P)^{-1} \bar{\epsilon}_k,
\end{equation}
\begin{equation}
d\theta_k =  \phi_k \cdot \min(\eta_k, \sqrt{C} / \norm{\phi_k}).
\end{equation}
}

\subsection{MinSR with Momentum \label{sec:minsr+m}} 
We motivate SPRING primarily via its relation to the Kaczmarz method. An alternative viewpoint is to see SPRING as a modification of a momentum method with an added projection step to remove the component of the momentum that conflicts with the current subproblem $\bar{O}_k d\theta = \bar{\epsilon}$. \rev{While we do not endorse this perspective,} it is natural to wonder how SPRING would compare to MinSR with na\"ive momentum, which we refer to as MinSR+M. 

\revB{Concretely, MinSR+M iteratively updates a momentum direction $\phi_k$ by taking a convex combination of the previous value $\phi_{k-1}$ and the current MinSR solution. Just like in SPRING, we assume an imaginary time step of $\delta \tau =1$ when we calculate $\bar{\epsilon}_k$ and instead control the step size using a learning rate parameter $\eta_k$ and a norm constraint. The method is thus described by the formulas:
\begin{equation}
\phi_k = (1-\mu) \bar{O}_k^T (\bar{O}_k \bar{O}_k^T + \lambda I + P)^{-1} \bar{\epsilon}_k + \mu \phi_{k-1},
\end{equation}
\begin{equation}
d\theta_k =  \phi_k \cdot \min(\eta_k, \sqrt{C} / \norm{\phi_k}).
\end{equation}}

\rev{As in SPRING, the $\mu$ parameter acts as a decay factor on the previous parameter update. However, MinSR+M differs significantly from SPRING in that no regularized projector is applied to the previous update. It also differs in that the MinSR solution is scaled by a factor of $(1-\mu)$ to achieve a convex combination, which means $\mu$ must be chosen to be strictly less than one. See Section \ref{sec:hyper} and Figure \ref{fig:momentum} for further discussion of the choice of $\mu$ for MinSR+M.} We demonstrate in our numerical experiments that SPRING outperforms MinSR+M substantially. \rev{This supports the viewpoint that SPRING is not simply a momentum method, but is rather a way of improving the quality of the parameter updates as approximate solutions to the SR equation.}

\section{Results}

We now turn to our numerical results. Our main finding is that \projsr{} outperforms KFAC, MinSR, and MinSR+M for several small atoms and molecules. We present results for the carbon, nitrogen, and oxygen atoms in Section \ref{sec:atoms} and for the $N_2$ \revB{and CO molecules} in Section \ref{sec:N2}. \revB{We summarize the final energies attained in our main experiments in Table \ref{tab:refE}}. We use VMCNet \cite{vmcnet2021github} to run all our experiments, which is a neural network VMC framework based on JAX \cite{jax2018github}. For all of our experiments, we use a standard FermiNet architecture with 16 dense determinants; see Table \ref{tab:ferminet_arch} for details.

For our training phase we use one hundred thousand optimization iterations with one thousand MCMC samples per iteration. For our MCMC proposals we use Gaussian all-electrons moves with a step-size that is tuned dynamically to maintain an acceptance ratio of approximately 50\%.  We take ten MCMC steps between optimization iterations to reduce correlations between the samples, and we use mean--centered local energy clipping to improve the stability of the optimization. We use a decaying learning rate with a decay rate of $r = 1e-4$ and learning rate schedule $\eta_k = \frac{\eta}{1 + rk}$. \revB{For MinSR, MINSR+M, and SPRING, we assume $\delta \tau=1$ and apply the learning rate $\eta_k$ as described in Section \ref{sec:meth}.} For more details on our VMC settings, see Table \ref{tab:vmc_hyp}. To obtain final energies we run a separate inference phase with the parameters from the last iteration of each optimization run; see Table \ref{tab:inf} for details. 

To compare the methods fairly, we do not start them from fully random initializations. Rather, for each system in question, we first run a short preliminary optimization and save the result to a checkpoint.  The rest of our experiments then load the starting parameters from the end of the preliminary optimization phase, switch out the optimizer and hyperparameters to the desired settings, and run one hundred thousand further optimization iterations. This procedure removes the chaotic early optimization stage as a factor from the comparison between the different methods, ensuring that no method is unfairly advantaged by randomly having a favorable start. Our preliminary optimization is different from the  Hartree-Fock pretraining procedure (see \cite{pfau2020ab, von2022self} for example) because we optimize the variational energy from the beginning. However, our results should translate well to a setting with pretraining as a result of this preliminary optimization. See Table \ref{tab:prelim_hyp} for more details on the preliminary optimization phase. 

For all of our optimizers we use a Tikhonov regularization parameter of $\lambda=0.001$ and a norm constraint of $C=0.001$ unless indicated otherwise. For MinSR+M we use a momentum parameter of $\mu=0.9$, and for \projsr{} we use the regularization parameter $\mu=0.99$. In Section \ref{sec:hyper} we present several hyperparameter studies which demonstrate that these values provide a fair basis for comparing the methods. \revB{We also investigate the effect of the preliminary optimization phase and demonstrate that SPRING outperforms KFAC significantly  without  preliminary optimization.}

In all of our numerical results we report energy errors relative to the benchmark energies listed in Table \ref{tab:refE}. For the energy and variance trajectories, we smooth out the curves by reporting averages over a sliding window of ten thousand iterations. We note that we are using a somewhat smaller network, fewer MCMC samples, and fewer optimization iterations than were used in most state-of-the-art calculations with KFAC; hence it should not be surprising that some of our results do not reach the same level of accuracy as those presented in the literature for comparable systems. Our results can be viewed as as proof-of-principle for the value of \projsr{} on small calculations, which will need to be extended to larger calculations by future works. 

\begin{table}
    \centering
    \footnotesize
    \begin{tabular}{c|c|c|c|c|c|c}
    System  & \revB{KFAC} & \revB{MinSR} & \revB{MinSR+M} & \revB{SPRING} & Benchmark & \revB{Benchmark source} \\
    \hline \hline 
    C & \revB{-37.8445} & \revB{-37.8445} & \revB{-37.8448}  & \revB{-37.8449} & -37.8450 & \cite{chakravorty1993ground}, Table XI\\
    N & \revB{-54.5877} & \revB{-54.5885} & \revB{-54.5889} & \revB{-54.5890} & -54.5892 & \cite{chakravorty1993ground}, Table XI \\ 
    O & \revB{-75.0643} & \revB{-75.0652} & \revB{-75.0659} & \revB{-75.0668} &  -75.0673 & \cite{chakravorty1993ground}, Table XI \\
    N$_2$, equilibrium & \revB{-109.5301} & \revB{-109.5268} & \revB{-108.5294} & \revB{-109.5322} &-109.5423  & \cite{filippi1996multiconfiguration}, Table II \\
    N$_2$, 4.0 Bohr & \revB{-109.1862} & \revB{-109.1794} & \revB{-109.1829} & \revB{-109.1906} &-109.2021  & \cite{le2006accurate}, MLR$_4(6,8)$ \\
    \revB{CO} & \revB{-113.3140} & \revB{-113.3092} & \revB{-113.3117} & \revB{-113.3169} & \revB{-113.3255} & \revB{\cite{pfau2020ab}, Table II} \\
    \end{tabular}
    \caption{\revB{Summary of final energies attained with tuned learning rates for all systems studied. Energies are reported in Hartrees, to four decimal places.}}
    \label{tab:refE}
\end{table}

\subsection{Small atoms \label{sec:atoms}}
\begin{figure*}
    \centering
    \begin{subfigure}[b]{0.475\textwidth}
        \centering
        \includegraphics[width=\textwidth]{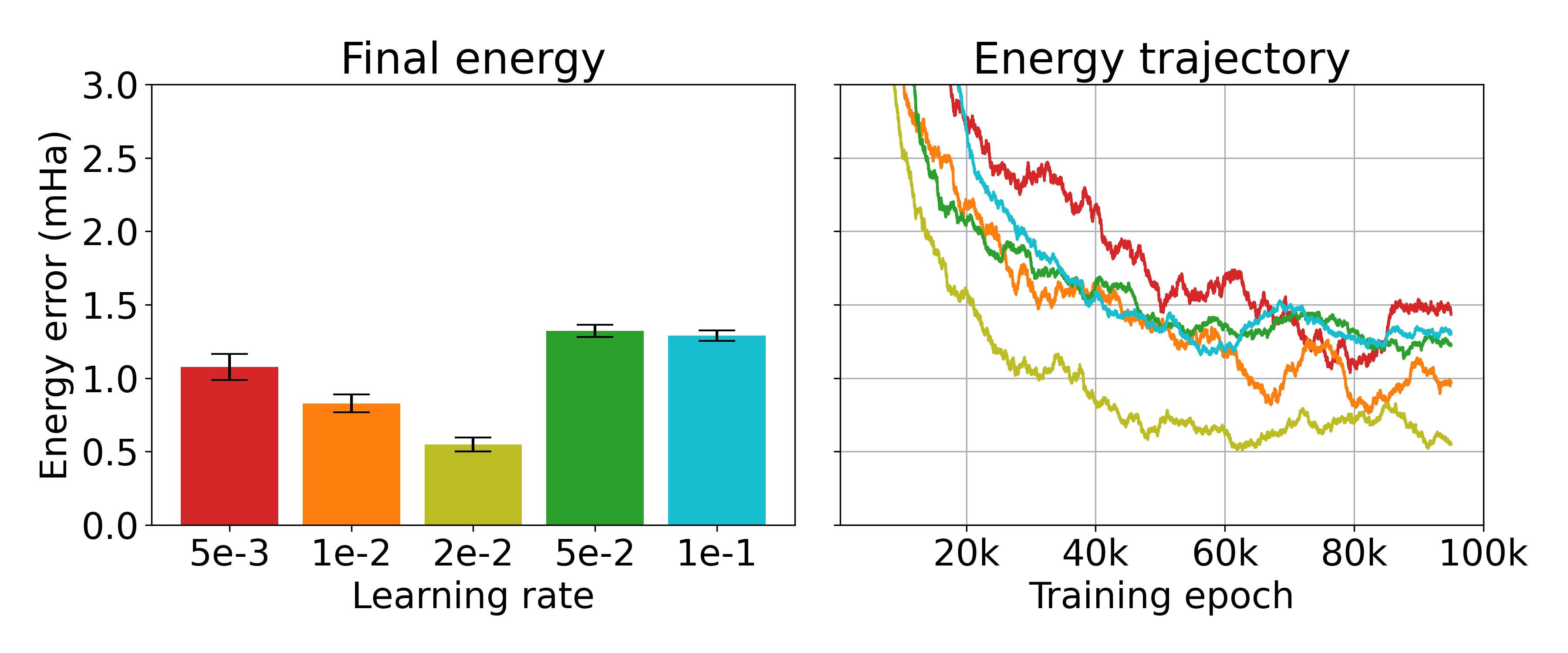} 
        \caption{KFAC}
    \end{subfigure}
    \hfill
    \begin{subfigure}[b]{0.475\textwidth}  
        \centering
        \includegraphics[width=\textwidth]{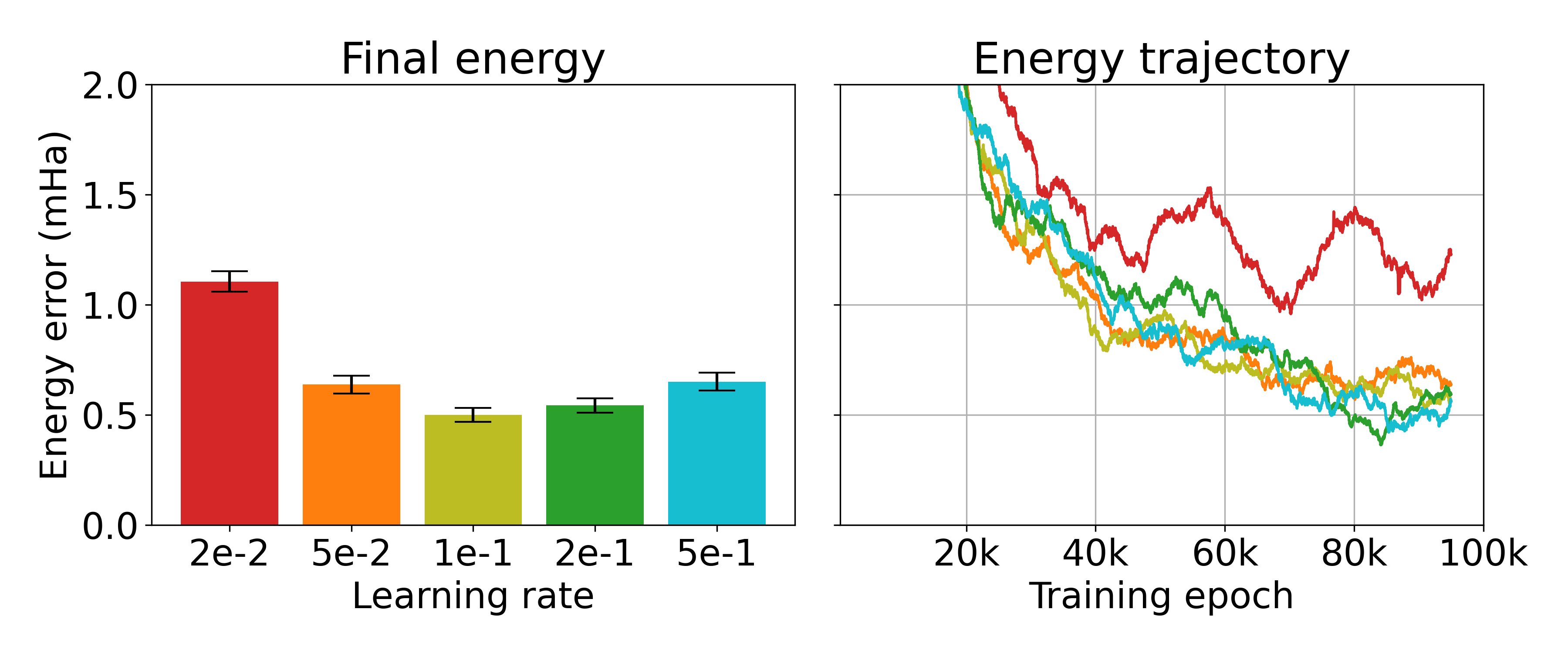}  
        \caption{MinSR}
    \end{subfigure}
    \vskip\baselineskip
    \begin{subfigure}[b]{0.475\textwidth}   
        \centering
        \includegraphics[width=\textwidth]{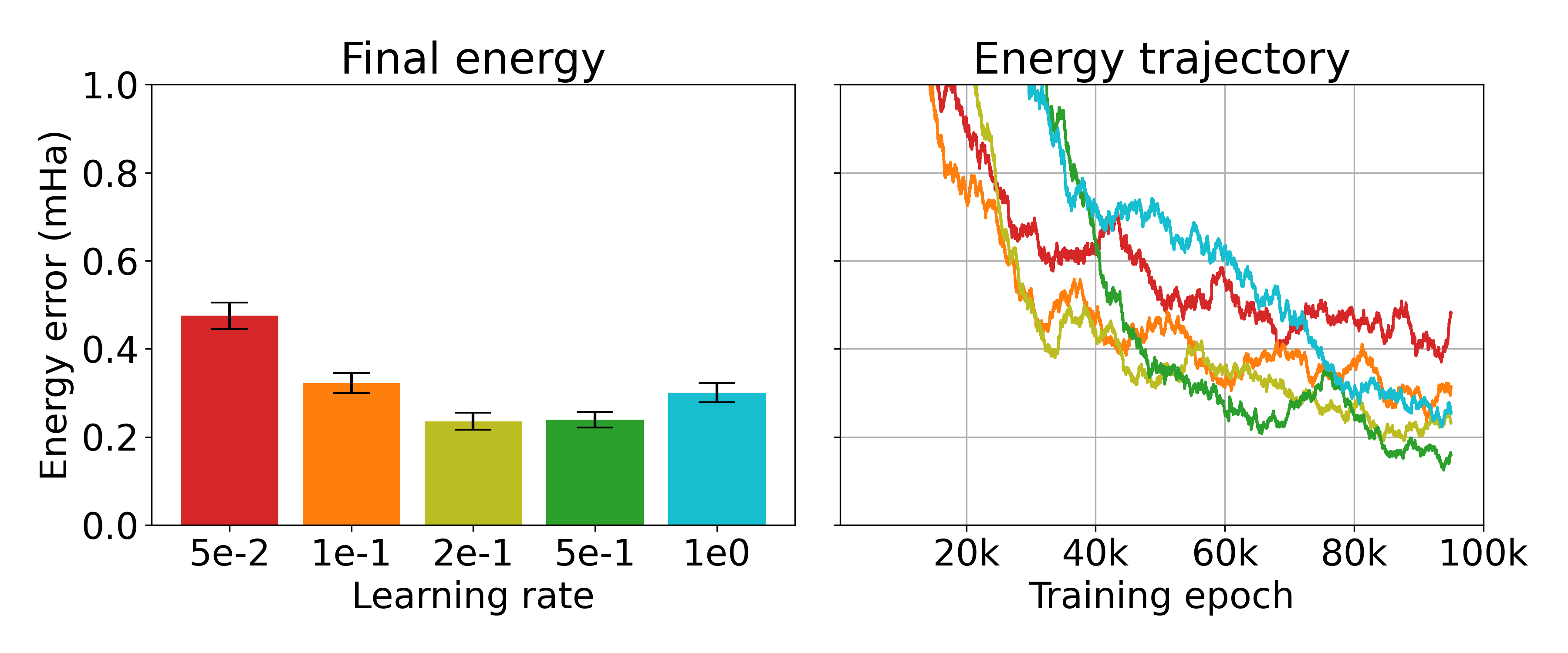}  
        \caption{MinSR+M}
    \end{subfigure}
    \hfill
    \begin{subfigure}[b]{0.475\textwidth}   
        \centering
        \includegraphics[width=\textwidth]{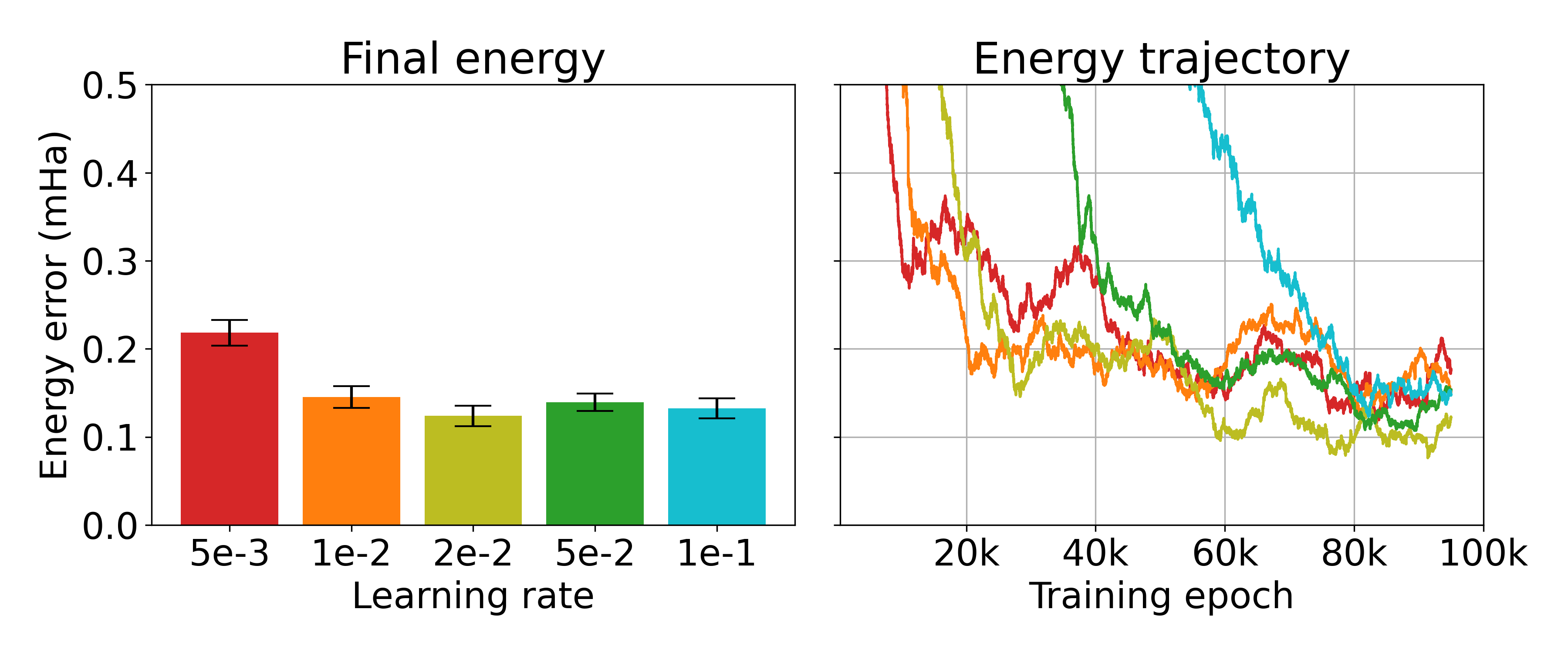}
        \caption{SPRING}
    \end{subfigure}
    \caption{Learning rate sweeps on the carbon atom with four different optimizers.}
   \label{fig:C_lr_constraint}
\end{figure*}

We first present results for KFAC, MinSR, MinSR+M, and \projsr{} on the carbon, nitrogen, and oxygen atoms. We find that it is critical to turn the learning rate of each method in order to achieve the best performance and obtain a fair comparison. Thus, for each method, we test a number of learning rates on the carbon atom, with results presented in Figure \ref{fig:C_lr_constraint}. Our experiments lead us to choose a learning rate of $\eta = 0.02$ for KFAC, $\eta = 0.1$ for MinSR, $\eta=0.2$ for MinSR+M, and $\eta = 0.02$ for \projsr{}. 

\begin{figure*}
    \centering
    \begin{subfigure}[b]{\textwidth}
        \centering
        \includegraphics[width=\textwidth]{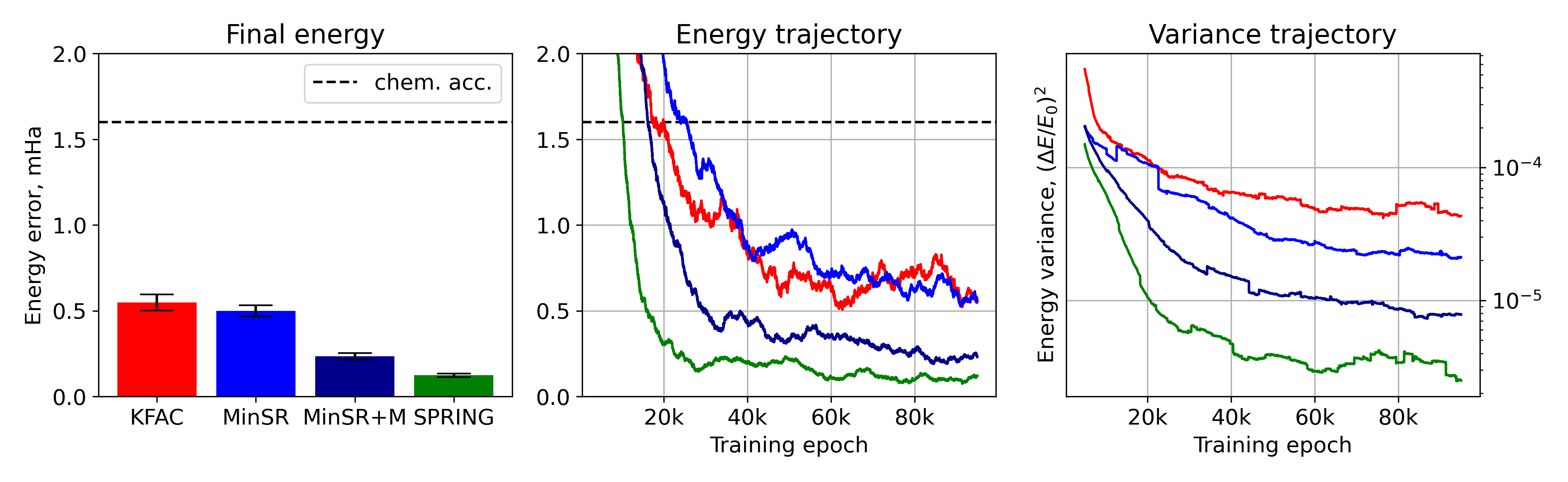} 
        \caption{Carbon atom.}
    \end{subfigure}
    \vskip\baselineskip
    \begin{subfigure}[b]{\textwidth}  
        \centering
        \includegraphics[width=\textwidth]{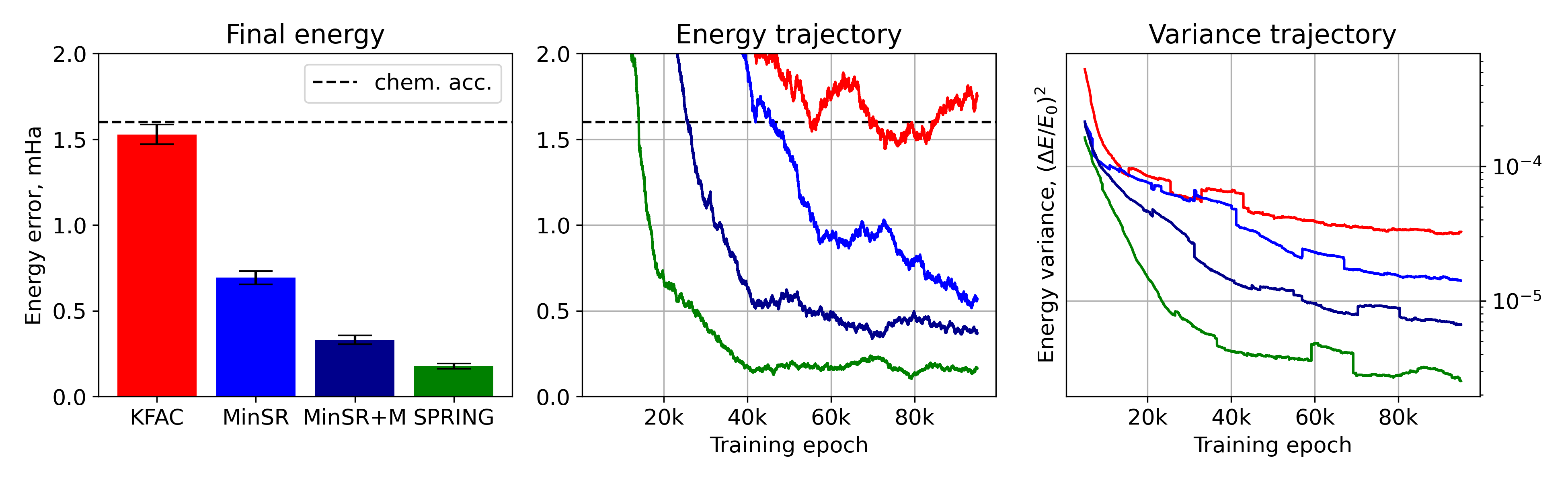}  
        \caption{Nitrogen atom.}
    \end{subfigure}
    \vskip\baselineskip
    \begin{subfigure}[b]{\textwidth}   
        \centering
        \includegraphics[width=\textwidth]{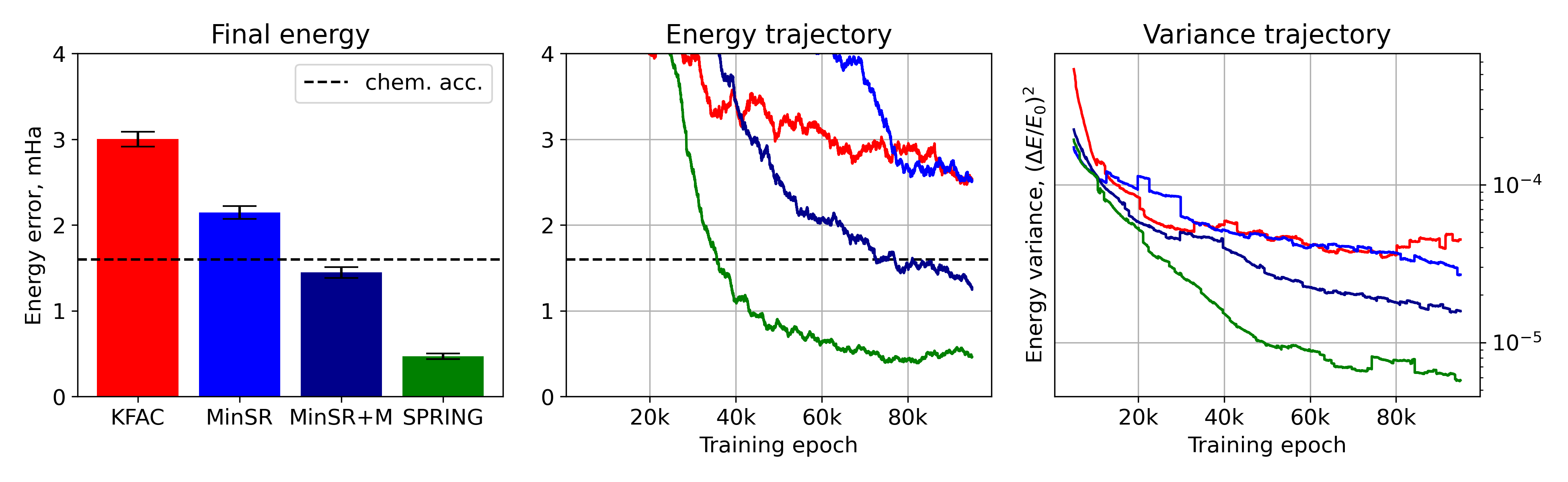}  
        \caption{Oxygen atom}
    \end{subfigure}
    \caption{Comparison of methods on three small atoms, with learning rates tuned on the carbon atom.}
   \label{fig:atom_results}
\end{figure*}

After tuning the learning rates on the carbon atom, we use the selected learning rates to compare the methods on carbon, nitrogen, and oxygen. The results with optimized learning rates are shown in Figure \ref{fig:atom_results}. On the carbon atom, we find that \projsr{} reaches chemical accuracy significantly faster than the other methods. Relative to KFAC and MinSR, its final energy error is about a factor of four lower, and its final local energy variance that is about an order of magnitude lower. On nitrogen and oxygen, \projsr{} continues to outperform the other methods. In the case of the oxygen atom, \projsr{} reaches chemical accuracy after forty thousand iterations, whereas MinSR+M takes eighty thousand iterations and the other methods never reach chemical accuracy.

\subsection{\revB{Molecules} \label{sec:N2}}
\begin{figure*}
    \centering
    \begin{subfigure}[b]{0.475\textwidth}
        \centering
        \includegraphics[width=\textwidth]{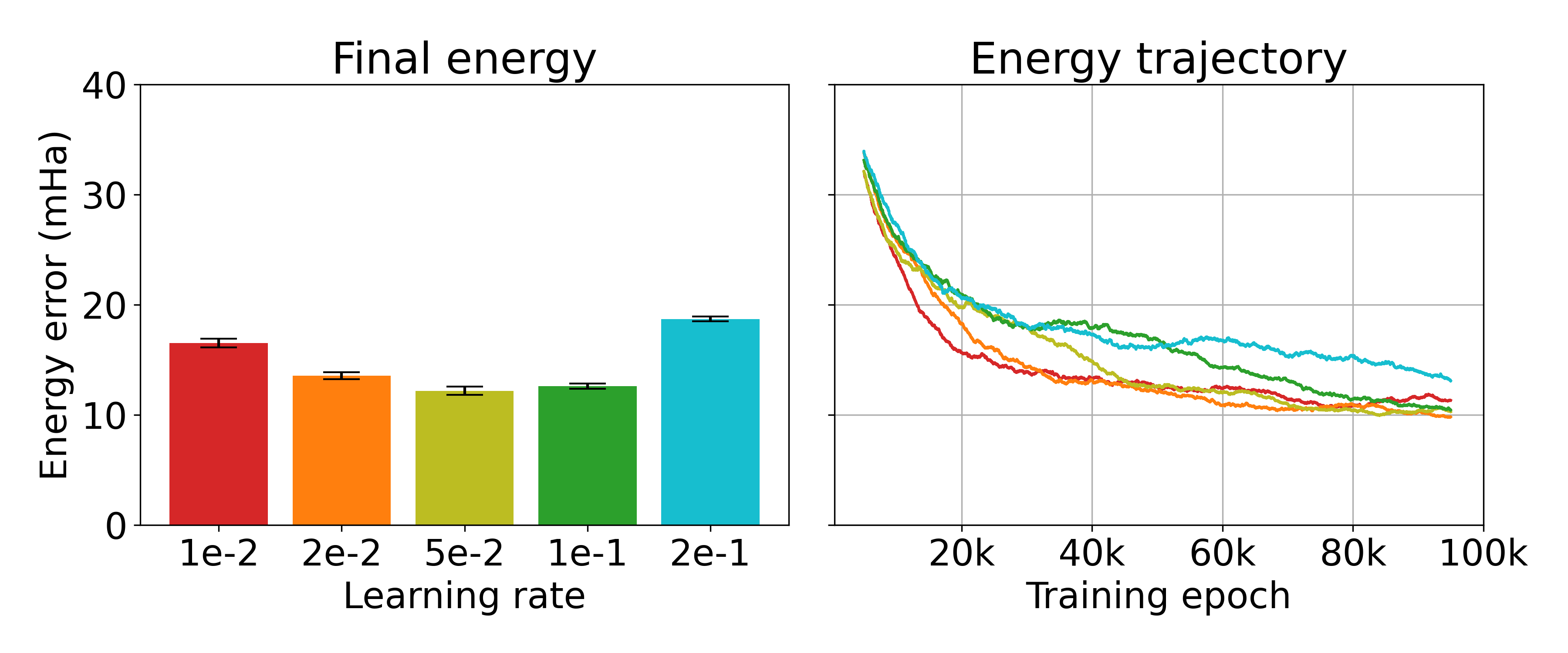} 
        \caption{KFAC}
    \end{subfigure}
    \hfill
    \begin{subfigure}[b]{0.475\textwidth}  
        \centering
        \includegraphics[width=\textwidth]{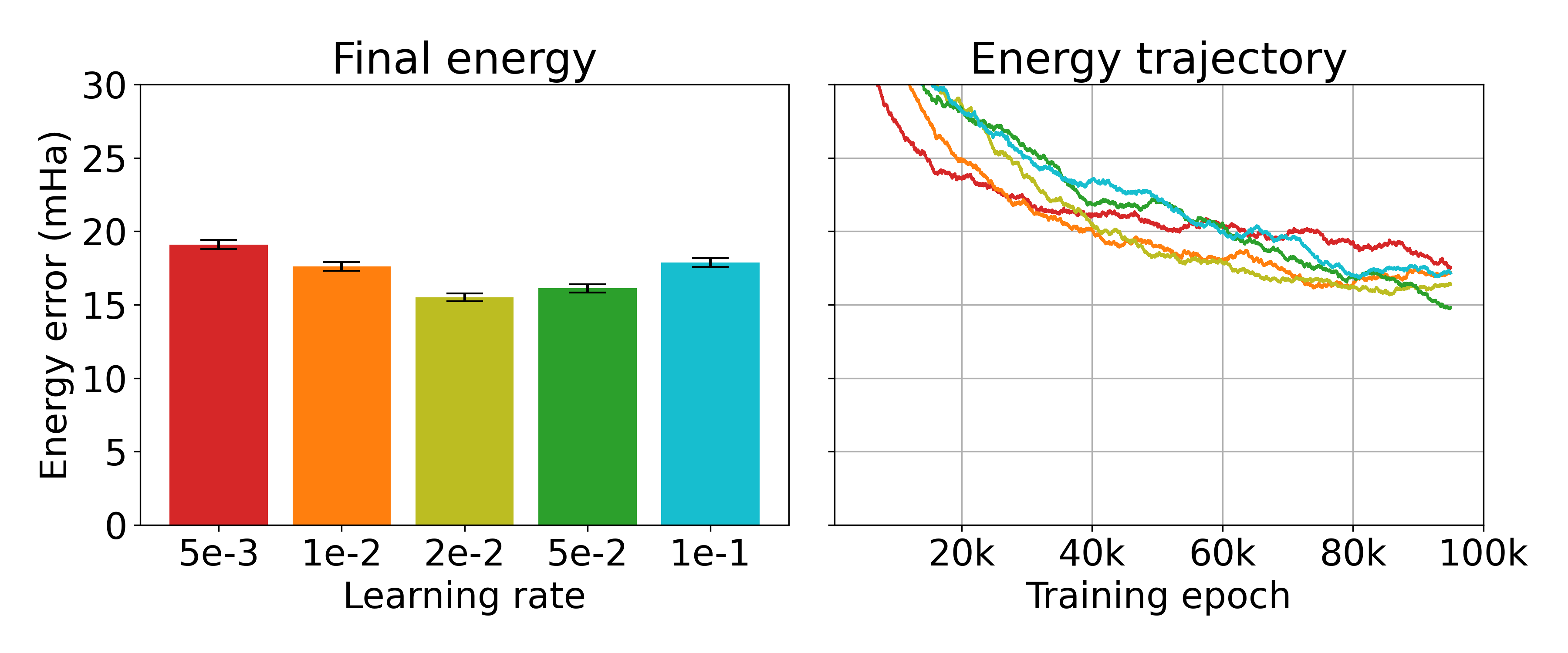}  
        \caption{MinSR}
    \end{subfigure}
    \vskip\baselineskip
    \begin{subfigure}[b]{0.475\textwidth}   
        \centering
        \includegraphics[width=\textwidth]{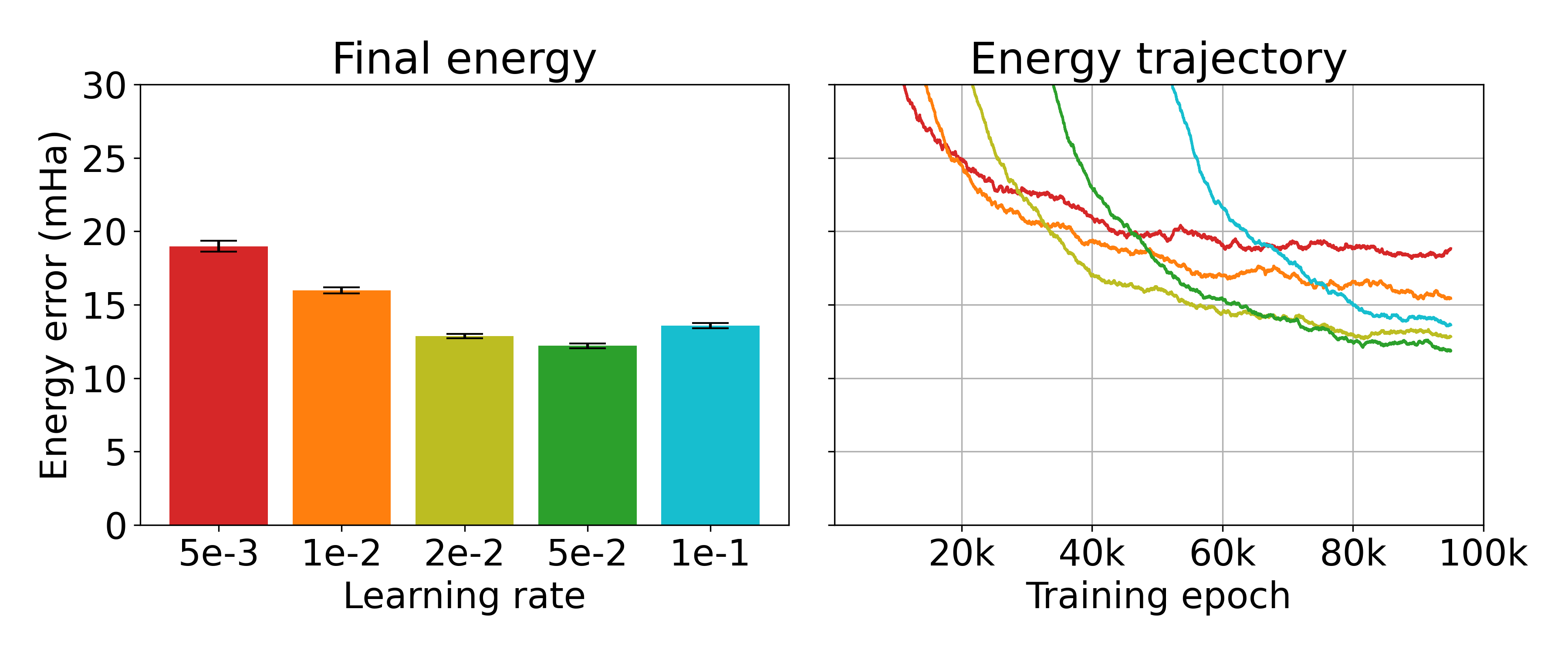}  
        \caption{MinSR+M}
    \end{subfigure}
    \hfill
    \begin{subfigure}[b]{0.475\textwidth}   
        \centering
        \includegraphics[width=\textwidth]{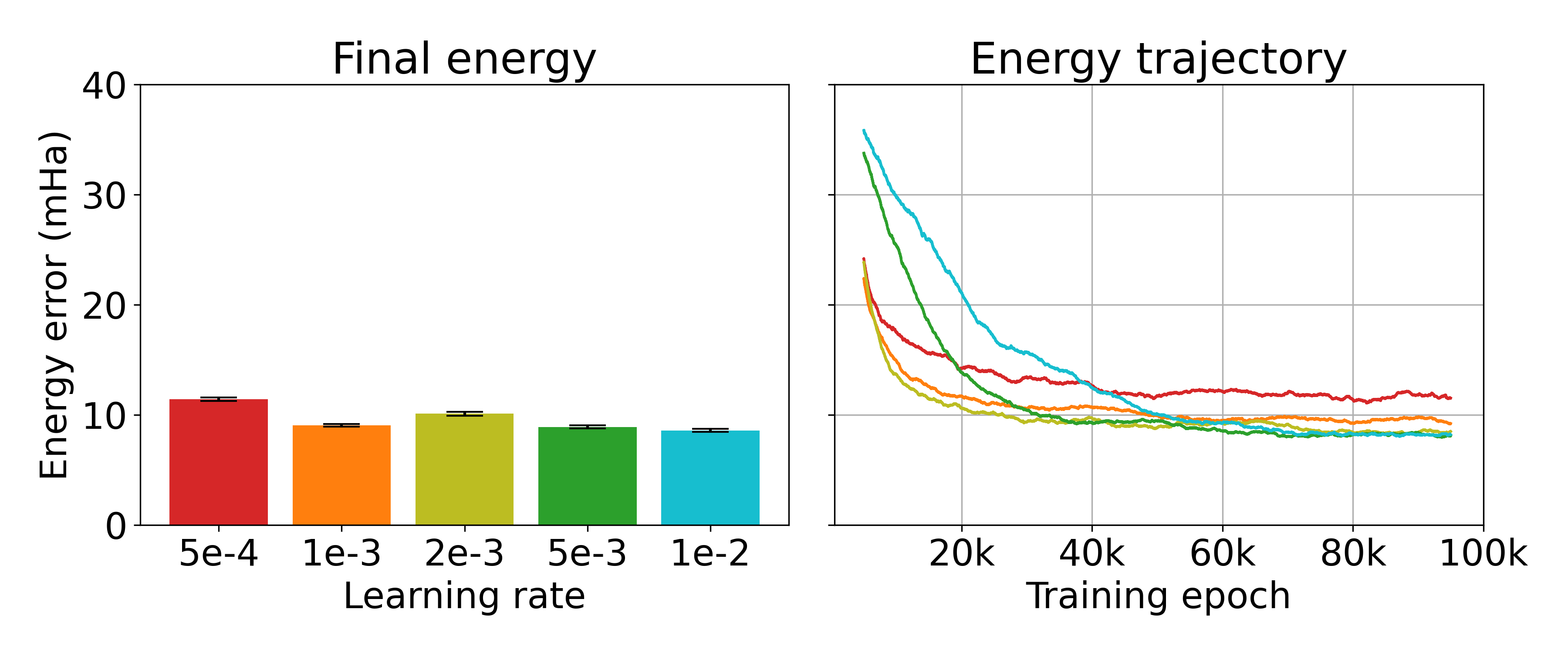}
        \caption{SPRING}
    \end{subfigure}
    \caption{Learning rate sweeps on the nitrogen molecule at equilibrium bond distance with four different optimizers.}
   \label{fig:N2_lr}
\end{figure*}

We next test KFAC, MinSR, MinSR+M, and \projsr{} on \revB{two somewhat larger systems: the N$_2$ and CO molecules. We begin by tuning the learning rates for all four methods on the N$_2$ molecule at an equilibrium bond distance of 2.016 Bohr}. We present the results of the learning rate sweeps in Figure \ref{fig:N2_lr}. Based on these results, we pick learning rates of $\eta = 0.05$ for KFAC, $\eta = 0.02$ for MinSR, $\eta=0.02$ for MinSR+M, and $\eta = 0.002$ for \projsr{}. These learning rates are not always chosen strictly for the best final energy since the final energy can vary due to statistical fluctuations in the parameters and statistical errors in the energy estimation. We instead look at the optimization trajectories and the final energies and pick the learning rate that appears to be optimal based on the combination of the two. It is worth noting that the optimal learning rates for MinSR+M and \projsr{} change by a factor of $10$ between the carbon atom and the N$_2$ molecule, whereas the optimal learning rate for MinSR changes by only a factor of $5$, and the optimal learning rate for KFAC changes by only a factor of $2$. The optimal learning rate may thus be more system-dependent for MinSR+M and \projsr{} compared to the other methods.

\begin{figure*}
    \centering
    \begin{subfigure}[b]{\textwidth}
        \centering
        \includegraphics[width=\textwidth]{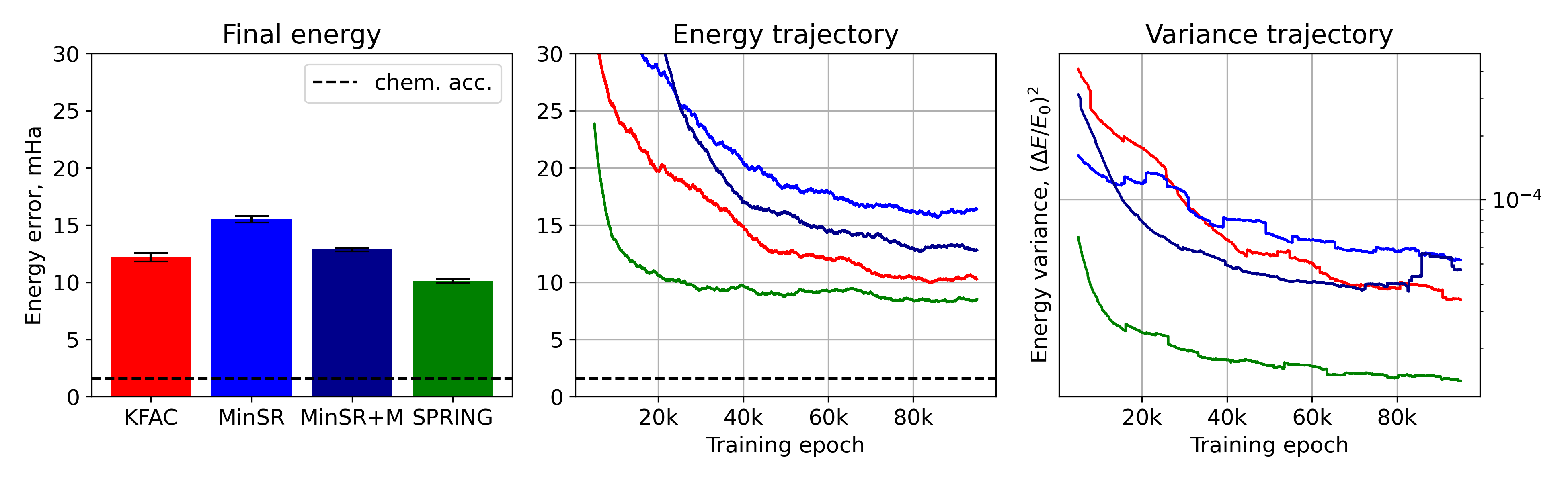} 
        \caption{Equilibrium configuration, bond distance 2.016 Bohr.}
    \end{subfigure}
    \vskip\baselineskip
    \begin{subfigure}[b]{\textwidth}  
        \centering
        \includegraphics[width=\textwidth]{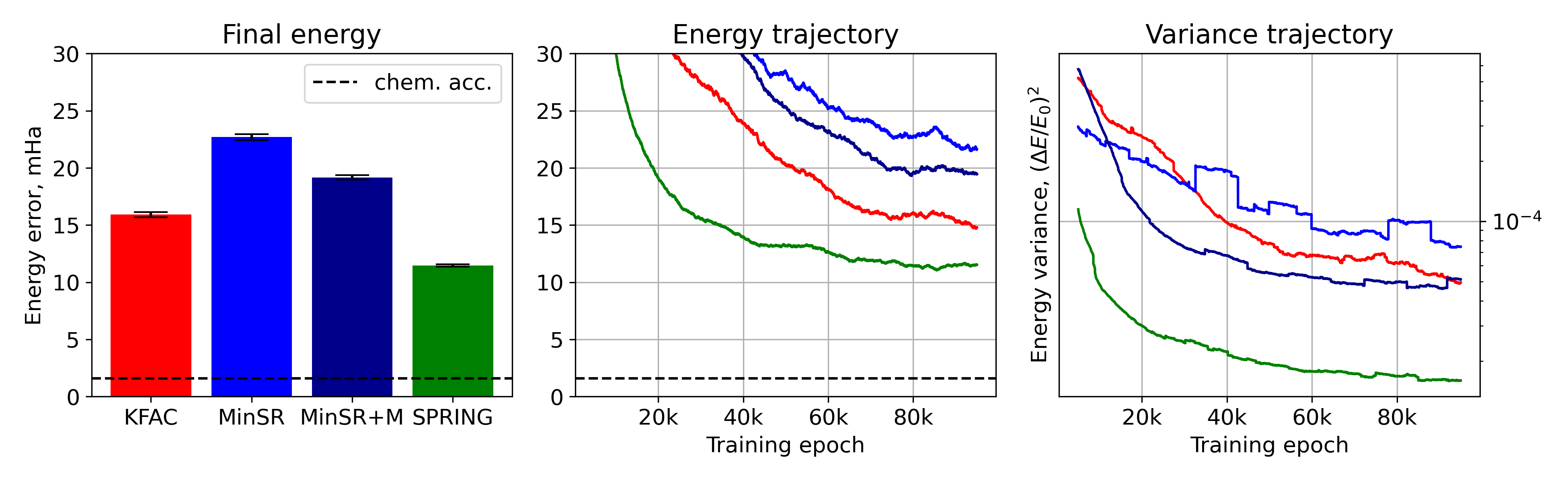}  
        \caption{Stretched configuration, bond distance 4.0 Bohr.}
    \end{subfigure}
    \caption{Comparison of methods on N$_2$ molecule at two bond distances, with learning rates tuned at equilibrium.}
   \label{fig:N2_results}
\end{figure*}

\begin{figure*}
    \centering
    \includegraphics[width=\textwidth]{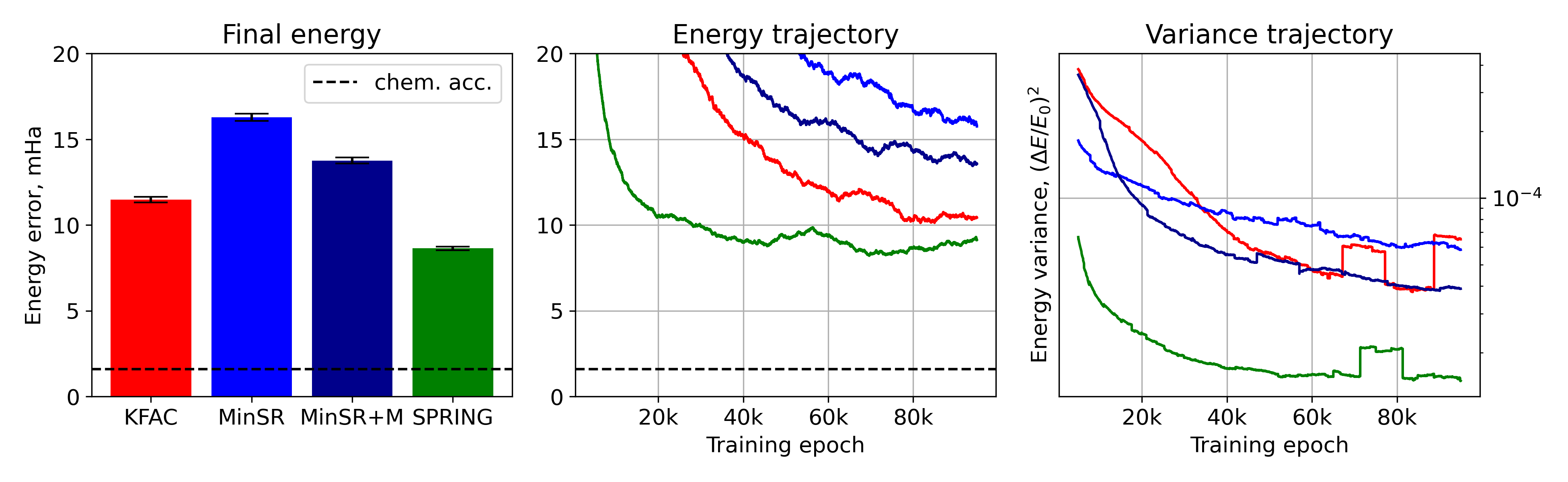} 
    \caption{\revB{Comparison of methods on CO molecule, with learning rates tuned on N$_2$.}}
   \label{fig:CO_results}
\end{figure*}

Using these optimized learning rates, we then compare the four methods \revB{on the N$_2$ molecule} both at equilibrium and at a stretched bond distance of 4.0 Bohr, \revB{and on the CO molecule at an equilibrium bond distance of 2.173 Bohr}. The results of these comparisons are presented in Figure \ref{fig:N2_results} \revB{and Figure \ref{fig:CO_results}}. For these systems none of the optimizers are able to reach chemical accuracy, which is expected due to the relatively small network, few MCMC samples, and few optimization iterations that we are using. Nonetheless, \projsr{} still displays a marked advantage over the other methods, converging much faster and to a lower energy and energy variance in \revB{all three} cases. The MinSR+M optimizer only improves upon MinSR marginally for \revB{these systems}, and it is outperformed by KFAC. This provides evidence that the SPRING algorithm is distinctly better than simply adding momentum on top of MinSR. 

\subsection{Hyperparameter studies \label{sec:hyper}}
In this section, we provide several experiments that demonstrate that our main results are robust to the choices of $C$ and $\mu$. \revB{We also provide evidence that the results are not significantly skewed by the preliminary optimization phase.} For simplicity we focus on the carbon atom for the purposes of this section.

\begin{figure*}
    \centering
    \begin{subfigure}[b]{0.475\textwidth}
        \centering
        \includegraphics[width=\textwidth]{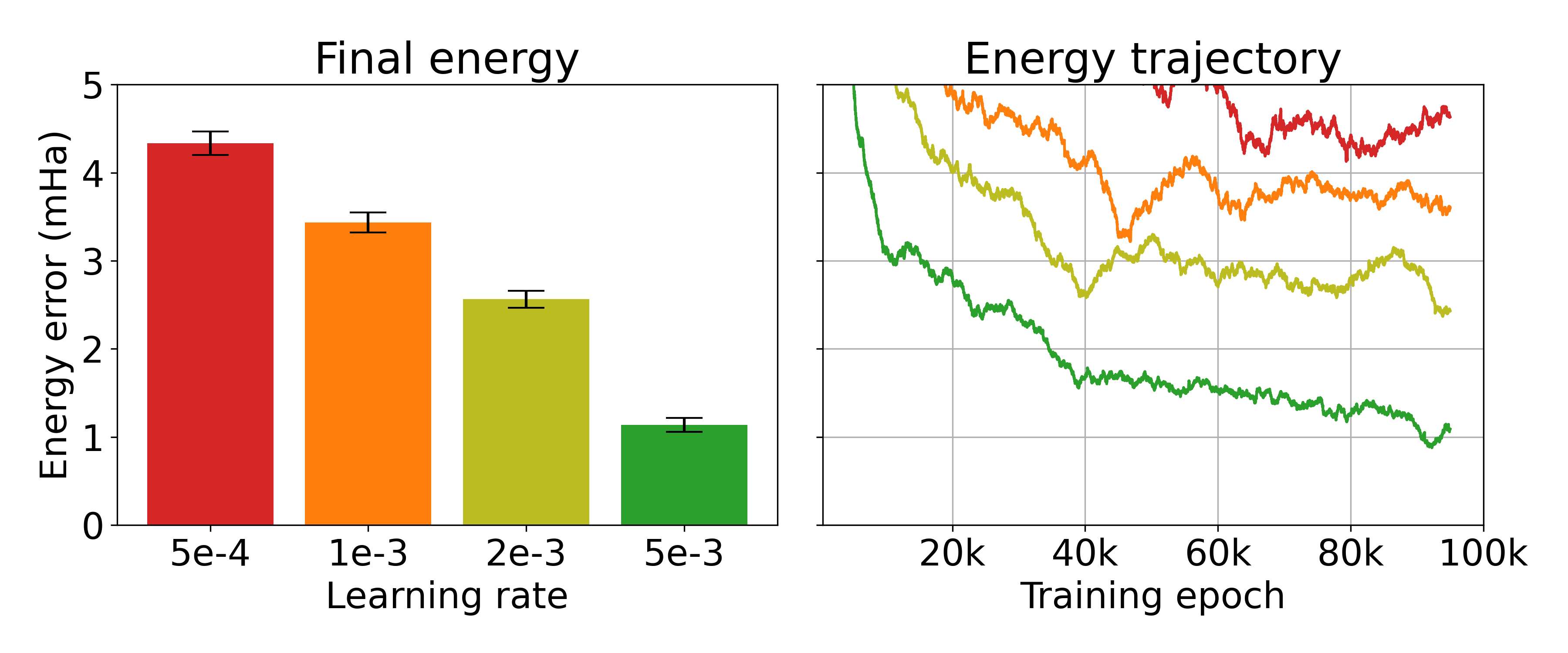} 
        \caption{KFAC}
    \end{subfigure}
    \hfill
    \begin{subfigure}[b]{0.475\textwidth}  
        \centering
        \includegraphics[width=\textwidth]{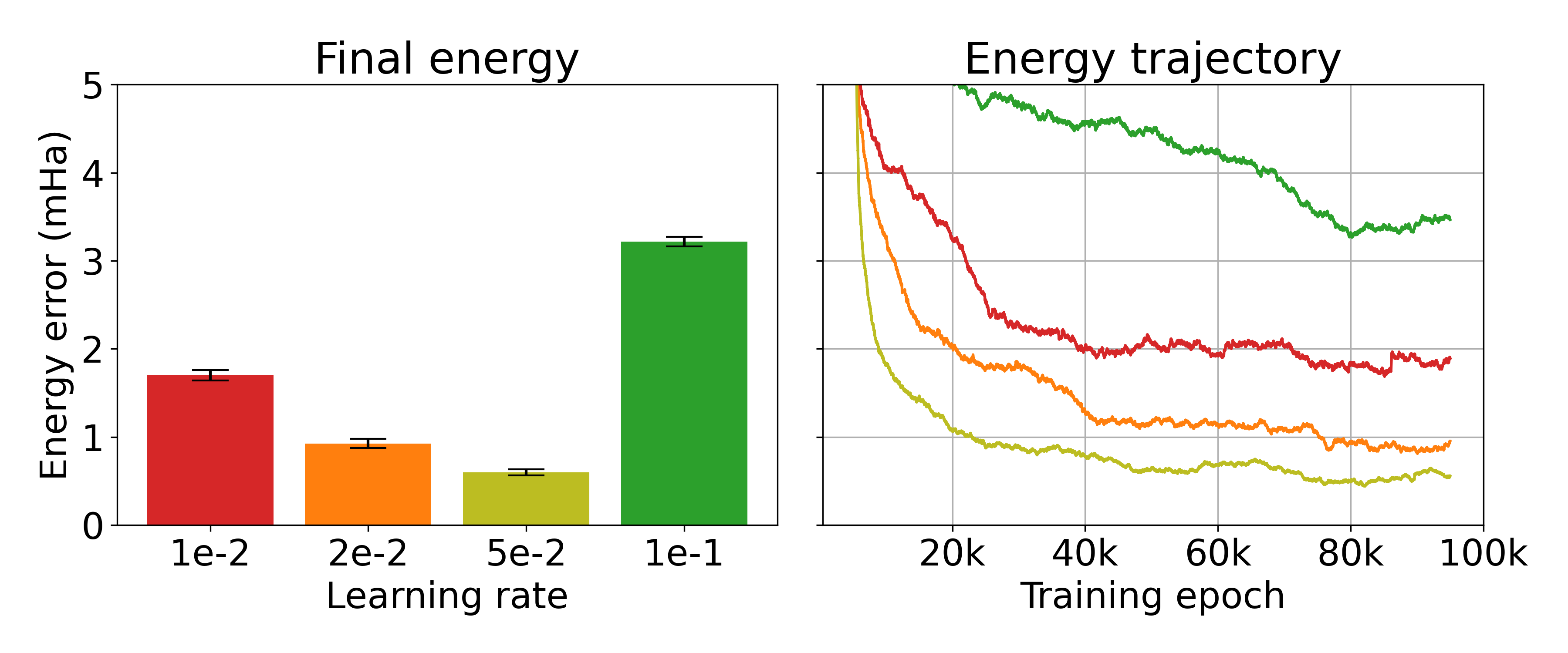}  
        \caption{MinSR}
    \end{subfigure}
    \vskip\baselineskip
    \begin{subfigure}[b]{0.475\textwidth}   
        \centering
        \includegraphics[width=\textwidth]{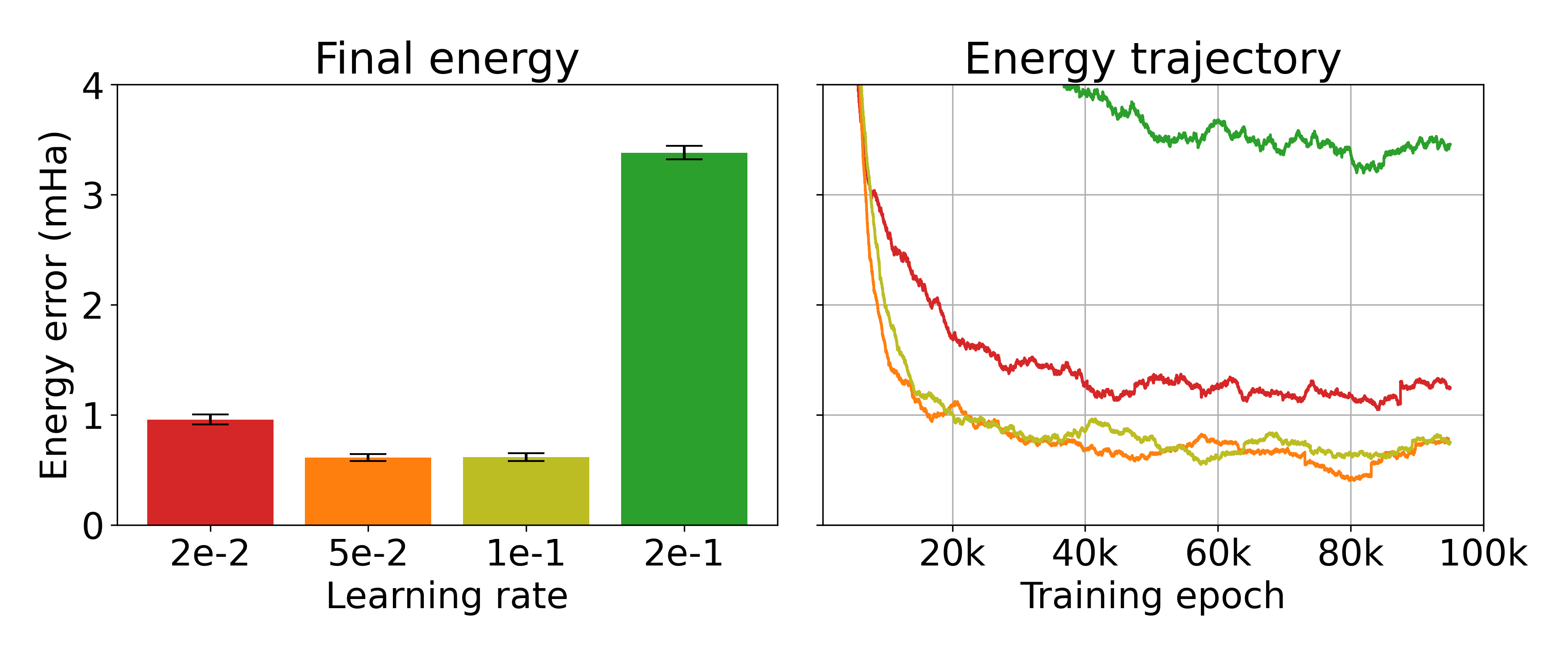}  
        \caption{MinSR+M}
    \end{subfigure}
    \hfill
    \begin{subfigure}[b]{0.475\textwidth}   
        \centering
        \includegraphics[width=\textwidth]{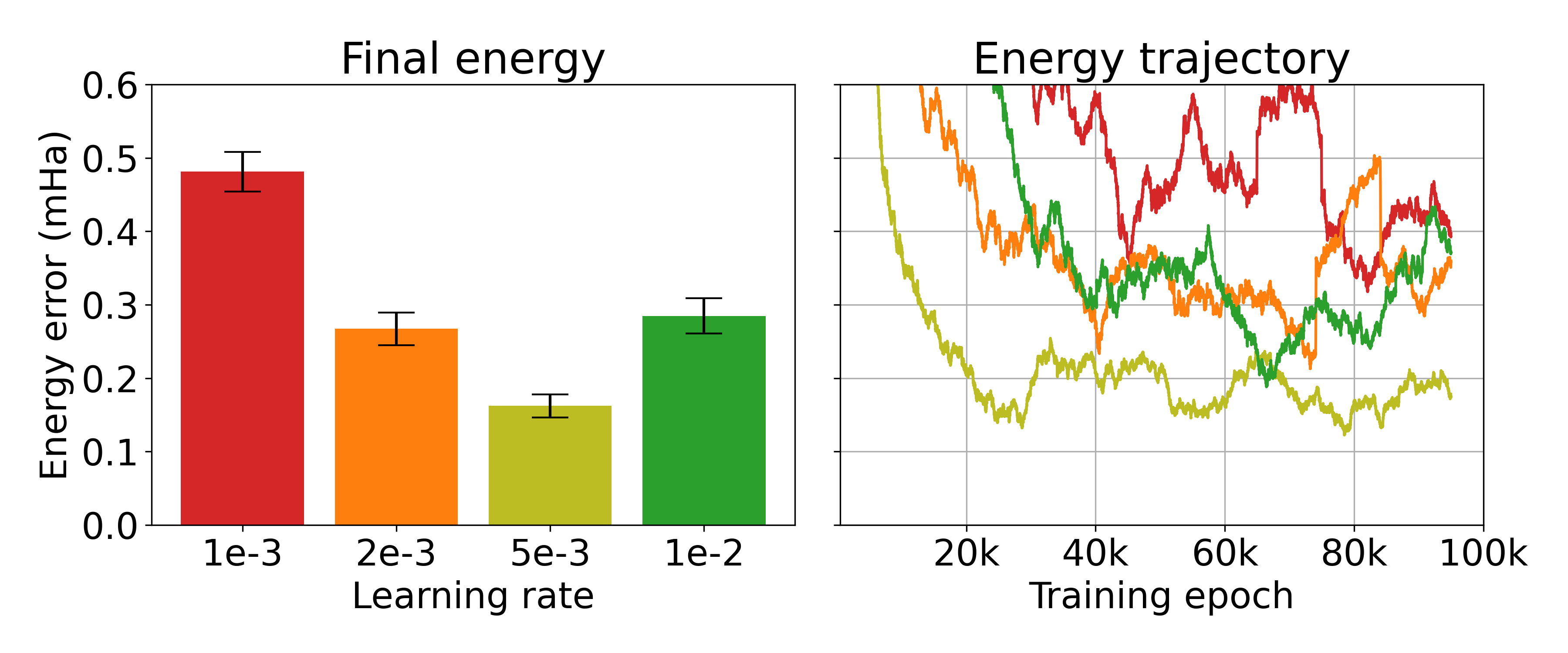}
        \caption{SPRING}
    \end{subfigure}
    \caption{Learning rate sweeps on the carbon atom without norm constraints with four different optimizers.}
   \label{fig:C_lr_noconstraint}
\end{figure*}

\begin{figure}
\centerline{
\includegraphics[width=\textwidth]{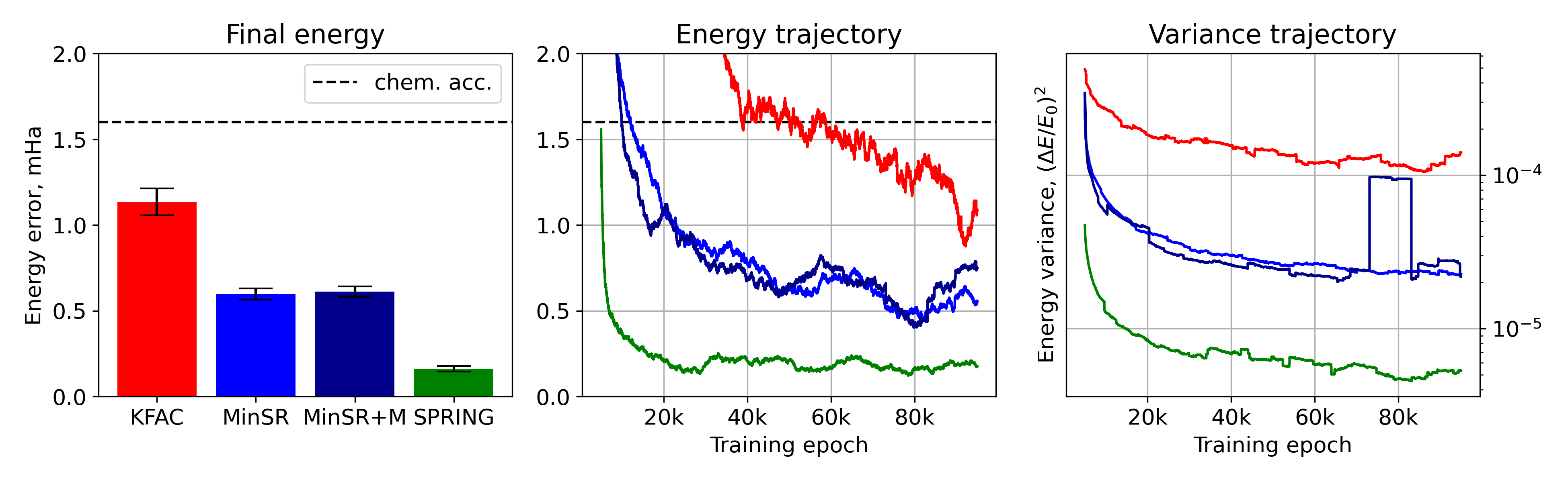}
}
\caption{Comparison of methods on carbon atom, with tuned learning rates and no norm constraint.}
\label{fig:C_nc_results}
\end{figure}

\begin{figure}
\centerline{
\includegraphics[width=0.4\textwidth]{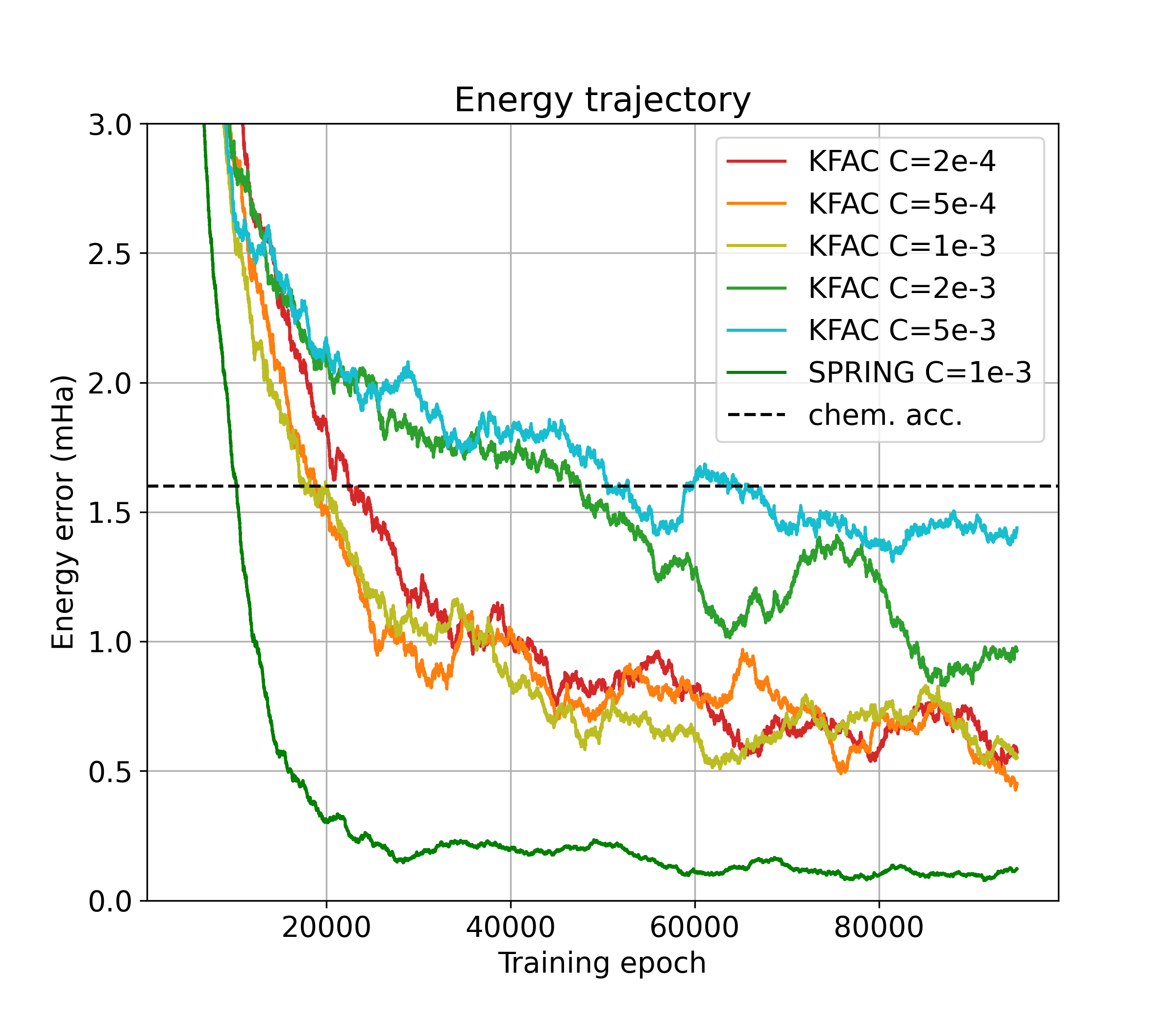}
}
\caption{Testing KFAC on the carbon atom with different values of the norm constraint $C$. No value performs better than the default value of $C=0.001$ and all settings are outperformed by \projsr{} with $C=0.001$. All runs use the optimized learning rate of $\eta=0.02$.}
\label{fig:C_kfac_C}
\end{figure}

We first show in Figure \ref{fig:C_lr_noconstraint} how the results of the learning rate experiments differ if we do not apply a norm constraint to any of the methods. In all four cases, the performance is much more sensitive to the learning rate when the constraint is turned off, and we are prevented from displaying results with larger learning rates because they produce completely unstable optimization trajectories. The combination of these factors means that without the norm constraints, it is much more difficult to find an effective learning rate, and our ability to transfer the learning rate from system to system is hindered. Nonetheless, even with the norm constraint off, the tuned version of \projsr{} significantly outperforms the tuned versions of the other methods, as shown in Figure \ref{fig:C_nc_results}. In Figure \ref{fig:C_kfac_C}, we show that neither increasing nor decreasing the norm constraint used for KFAC enables it to compete with the performance of \projsr{} on the carbon atom. This provides further evidence that the advantage of \projsr{} is not related to the choice or the form of the norm constraint.

\begin{figure*}
    \centering
    \includegraphics[width=0.4\textwidth]{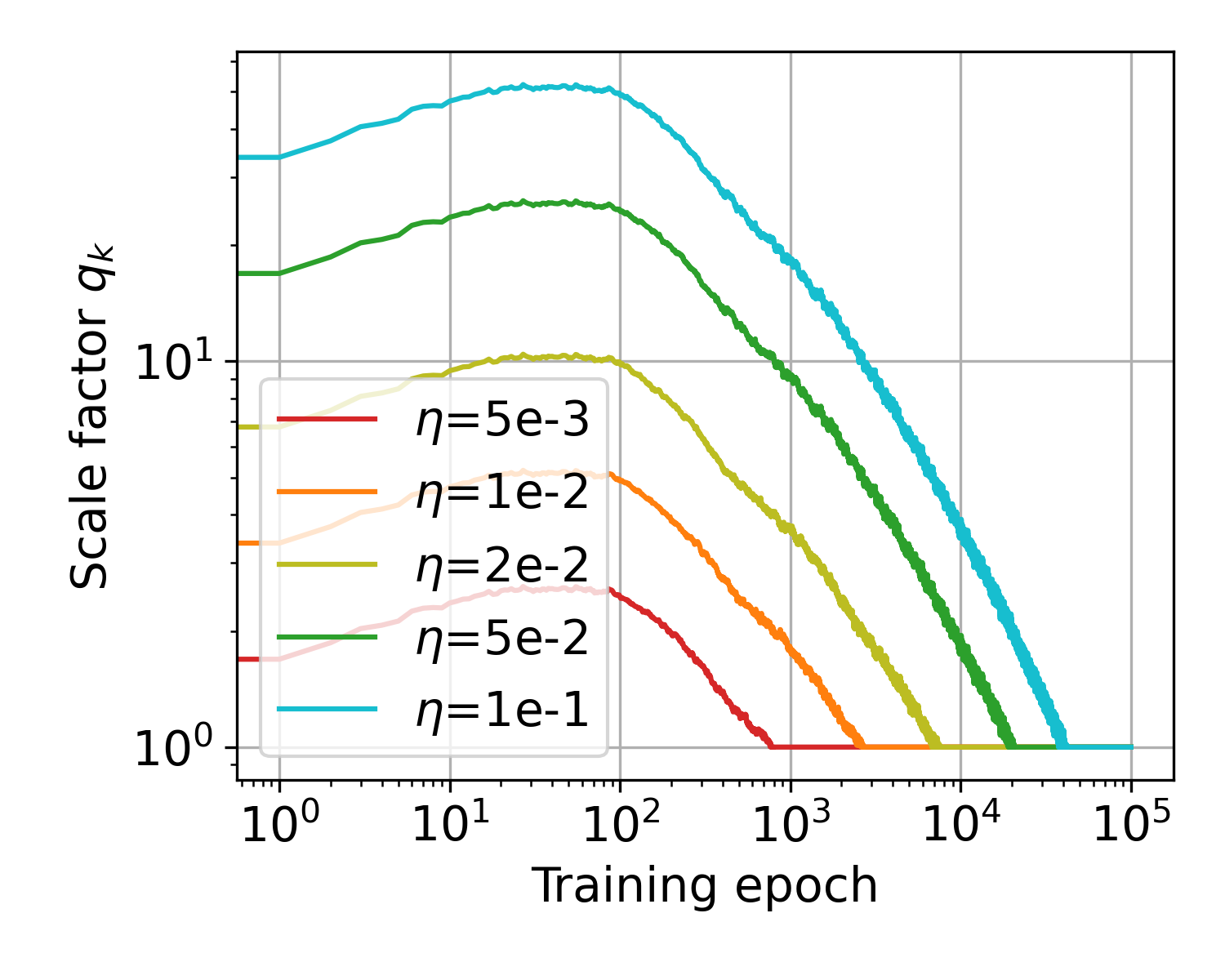} 
    \caption{\revB{Effect of the norm constraint on the SPRING parameter update during the learning rate sweep on the carbon atom. The quantity $q_k = \max(1, \norm{\eta_k \phi_k} / \sqrt{C})$ represents the extent to which the parameter update is scaled down as a result of the norm constraint.}}
   \label{fig:q}
\end{figure*}

\revB{
To better understand the impact of the norm constraint on SPRING, we return to the learning rate sweep for SPRING on the carbon atom and provide some additional data from these five optimization runs. In particular, we plot the scale factor $q_k = \max(1, \norm{\eta_k \phi_k} / \sqrt{C})$. For example, if $q_k=5$ then this means that at iteration $k$ the norm constraint resulted in a scaling down of the parameter update by a factor of $5$, and if $q_k=1$ then the norm constraint at iteration $k$ had no effect. We see in Figure \ref{fig:q} that $q_k$ tends to grow over the course of the first one hundred optimization iterations and then decay to $1$ as the optimization progresses. We can understand this in the following way: $q_k$ first rises as more optimization history is incorporated into SPRING; it then decays due to both the learning rate decay and the decay of the energy gradient. As should be expected, with a larger learning rate, $q_k$ reaches a larger peak and takes longer to decay to $1$.
}

\begin{figure*}
    \centering
    \begin{subfigure}[b]{0.475\textwidth}  
        \centering
        \includegraphics[width=\textwidth]{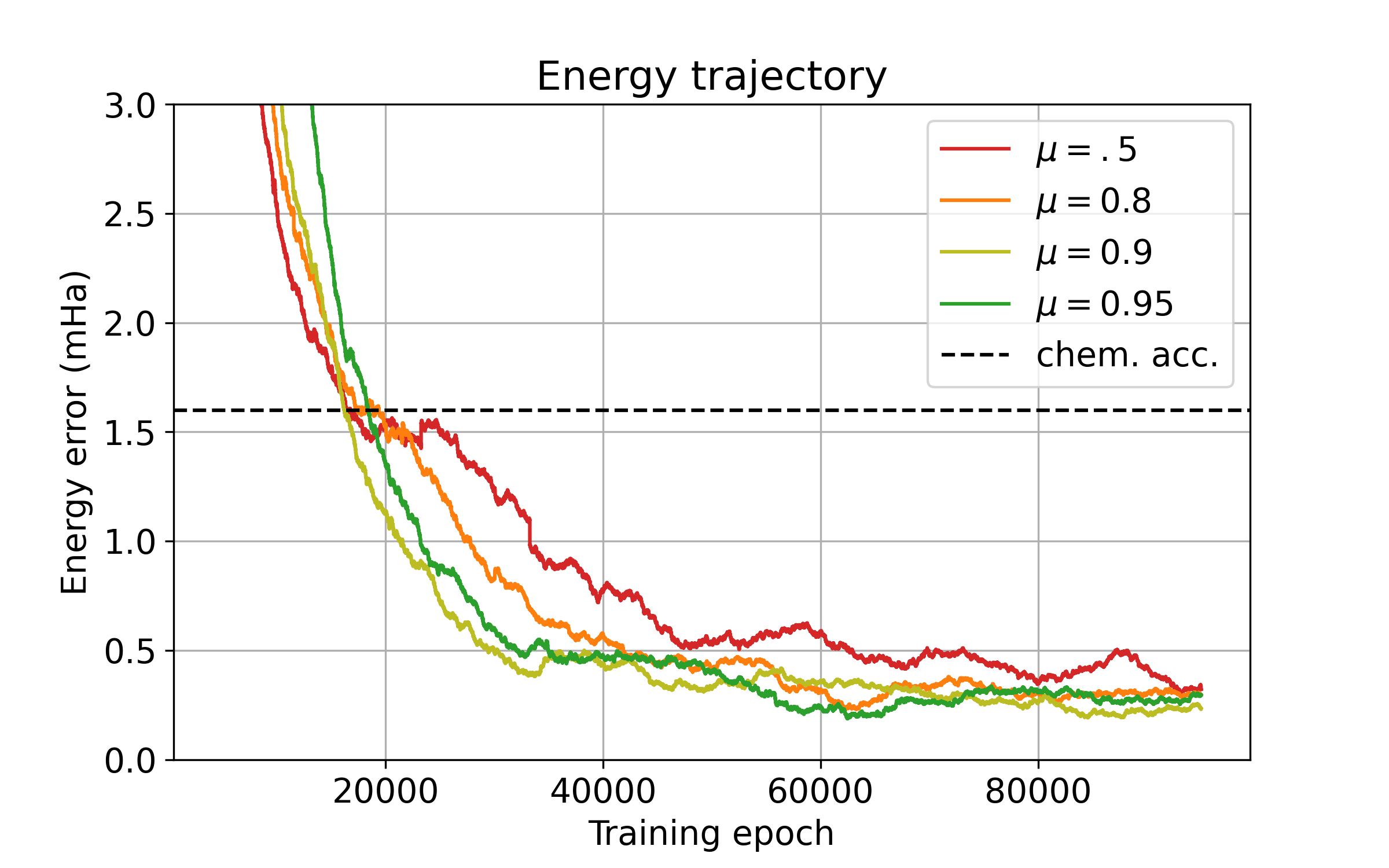}
        \caption{}
        \label{fig:momentum}
    \end{subfigure}
    \hfill
    \begin{subfigure}[b]{0.475\textwidth}   
        \centering
        \includegraphics[width=\textwidth]{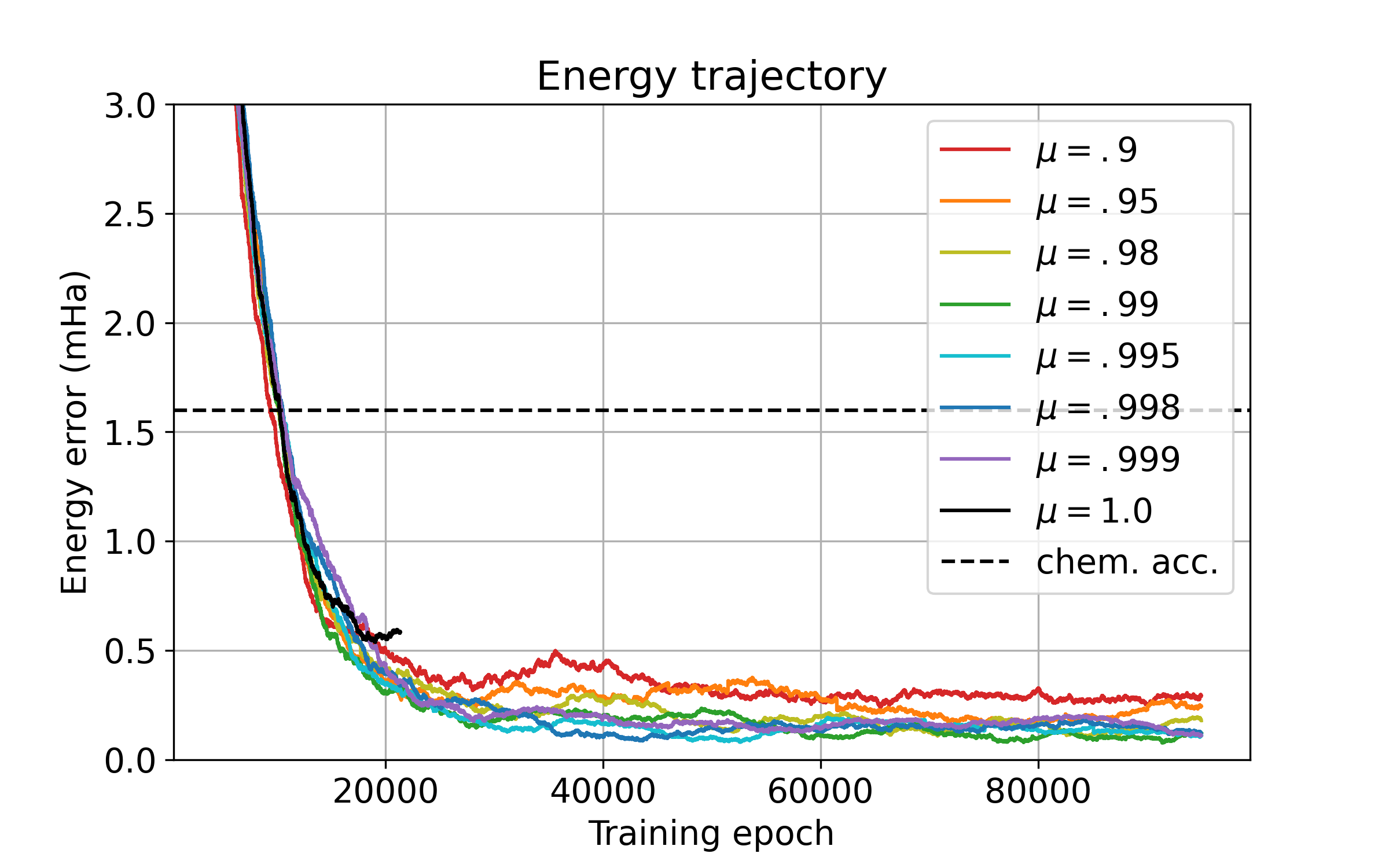}
        \caption{}
        \label{fig:mu}
    \end{subfigure}
    \caption{Hyperparameter studies for the parameter $\mu$ for both MinSR+M and SPRING. (a) MinSR+M with several values of the momentum parameter $\mu$, on the carbon atom. The learning rate is held fixed at the previously tuned value of $\eta=0.2$. (b) \projsr{} on the carbon atom with several values of the regularization parameter $\mu$. All values converge well except for the unregularized case $\mu=1.0$, shown in black, for which the optimization is unstable and encounters NaNs after approximately twenty thousand epochs. The learning rate is held fixed at the previously tuned value of $\eta=0.02$}
   \label{fig:hyp}
\end{figure*}

Next, we test the MinSR+M scheme with several different values of the momentum parameter $\mu$, using the previously optimized learning rate of $\eta=0.2$. For values larger than $\mu=0.95$ we find that the optimization is unstable and we do not report numerical results. We find that choosing $\mu$ anywhere between $0.8$ and $0.95$ produces approximately optimal results, which justifies our use of $\mu=0.9$ for our main experiments. Results are shown in Figure \ref{fig:momentum}. \revB{Similarly}, we test \projsr{} with several values of its regularization parameter $\mu$, again using the previously optimized learning rate of $\eta=0.02$. Results are shown in Figure \ref{fig:mu}. We find that choosing $\mu$ anywhere between $0.98$ and $0.999$ produces approximately optimal results, which justifies our choice of $\mu=0.99$ for our main experiments. Interestingly, if $\mu=1.0$ then the method becomes unstable. We do not yet have an explanation for the source of this instability.

\begin{figure*}
    \centering       
    \includegraphics[width=0.4\textwidth]{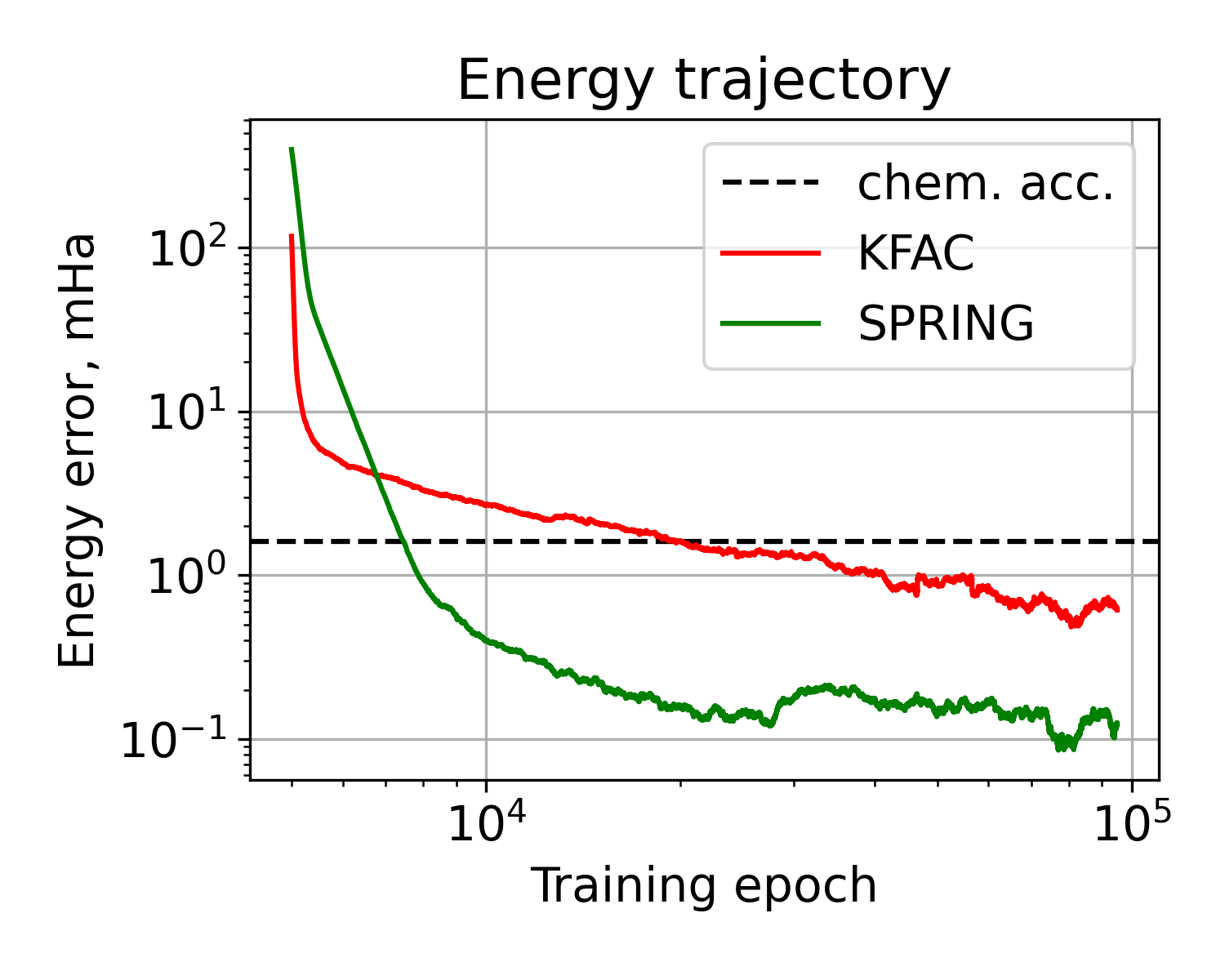}
    \caption{\revB{Comparison of SPRING versus KFAC on the carbon atom without preliminary optimization. We use the previously tuned learning rate of $\eta = 0.02$ for both KFAC and SPRING.}}
    \label{fig:cold}
\end{figure*}

\revB{Finally, we consider the effects of the preliminary optimization phase. Since we have used KFAC for the preliminary optimization in all cases, it is unclear thus far whether SPRING can perform well when run from a random initialization. To ameliorate this concern, we compare the performance of SPRING against KFAC without any preliminary optimization, with results in Figure \ref{fig:cold}. We find that KFAC outperforms SPRING in the very early stages of the optimization. However, SPRING overtakes KFAC after about seven thousand training iterations and reaches chemical accuracy in fewer than half as many iterations as KFAC. We conclude that SPRING is also effective  starting from a random initialization.}

\subsection{Computational Cost}
The computational cost of \projsr{} is essentially identical to that of MinSR, as the extra computations for \projsr{} are negligible in cost. The same applies to MinSR+M. For all three of these methods, the computational bottleneck lies in computing the matrix $T = \bar{O} \bar{O}^T$, which has asymptotic complexity $O(N_p \cdot N_s^2)$. This cost dominates the $O(N_s^3)$ cost of the Cholesky factorization.

Relative to KFAC, we expect the cost of \projsr{} to be highly dependent on the system size, the number of MCMC samples per iteration, and the number of GPUs used to parallelize the calculation, and we do not attempt to perform a systematic study along these lines. In our experiments, we use a single GeForce RTX 2080Ti GPU with 1000 MCMC walkers. With these particular settings, we find that \projsr{} results in VMC iterations that are about $50\%$ slower than KFAC for our smaller systems such as the carbon atom. As system size grows, the difference is reduced, and for the $N_2$ molecule the VMC iterations with \projsr{} are only about $5\%$ slower than with KFAC.

\section{Discussion}
Neural network wavefunctions represent a promising avenue towards highly accurate simulations of small but challenging molecular systems. The major bottleneck of applying such wavefunctions is the high cost associated with their optimization. In this work we introduce a new optimizer called \projsr{} to alleviate this bottleneck. SPRING combines ideas from the recently proposed MinSR optimizer \cite{minsr} with ideas from the randomized Kaczmarz method for solving overdetermined least-squares problems \cite{strohmer2009randomized}. By doing so, SPRING is able to utilize optimization history to improve upon MinSR in a principled way at essentially no extra cost. We test \projsr{} on several small atoms and molecules, comparing it against MinSR, MinSR with momentum, and KFAC. We find that SPRING consistently outperforms the alternatives across all tested systems. We hope that these findings will be extended to larger systems by future works. 

Due to several discrepancies between the VMC setting and the traditional Kaczmarz setting, we are not yet able to furnish a rigorous proof of convergence for the SPRING optimizer. One interesting direction for future research is to modify SPRING in a way that makes it possible to prove its convergence. For example, the exponentially convergent randomized Kaczmarz method of Strohmer and Vershynin requires that the rows are sampled with probability proportional to the square of their Euclidean norm \cite{strohmer2009randomized}. Furthermore, in the randomized block Kaczmarz method, Needell and Tropp find that for optimal performance it is critical to partition the constraints of the least-squares problem into well-conditioned blocks \cite{needell2014paved}. Such sampling schemes would represent a significant departure from traditional VMC methods that sample from the probability density of the wavefunction, but they could lead to rigorous convergence guarantees, better performance, or both. There are also variants of the Kaczmarz method such as the randomized extended Kaczmarz algorithm of Zouzias and Freris \cite{zouzias2013randomized} that can converge to the solution of inconsistent least-squares problems. This technique cannot be directly applied to the VMC setting since it requires sampling the columns as well as the rows of the system. However, it could serve as inspiration for further development of the algorithm.

We also draw a connection between the MinSR and \projsr{} methods for optimizing neural network wavefunctions and the efficient subsampled natural gradient descent method of Ren and Goldfarb \cite{ren2019efficient}. In particular, we show in \ref{app:NGD-SMW} that MinSR can be viewed as a simplified implementation of the method of Ren and Goldfarb which applies as long as the gradient of the loss function is a linear combination of the model gradients at the sampled points. This is quite a common scenario in machine learning, occurring for example in supervised learning with a mean-squared error loss function. In such a setting, \projsr{} can then be viewed as a potential improvement to existing subsampled natural gradient descent methods. We leave it to future works to determine whether SPRING can yield performance improvements for applications outside of VMC.

\section*{Acknowledgements}
This research used the Savio computational cluster resource provided by the Berkeley Research Computing program at the University of California, Berkeley (supported by the UC Berkeley Chancellor, Vice Chancellor for Research, and Chief Information Officer). This material is based upon work supported by the U.S. Department of
Energy, Office of Science, Office of Advanced Scientific Computing Research, Department of
Energy Computational Science Graduate Fellowship under Award Number(s) DE-SC0023112 (G.G.). This effort was supported by the SciAI Center, and funded by the Office of Naval Research (ONR), under Grant Number N00014-23-1-2729 (N.A.). LL is a Simons Investigator in Mathematics. We thank  Yixiao Chen, Zhiyan Ding, Yuehaw Khoo, Michael Lindsey, Eric Neuscamman, and Zaiwen Wen for their helpful discussions, \revC{and the anonymous reviewers for their valuable comments.}

\section*{Disclaimer}
This report was prepared as an account of work sponsored by an agency of the
United States Government. Neither the United States Government nor any agency thereof, nor
any of their employees, makes any warranty, express or implied, or assumes any legal liability
or responsibility for the accuracy, completeness, or usefulness of any information, apparatus,
product, or process disclosed, or represents that its use would not infringe privately owned
rights. Reference herein to any specific commercial product, process, or service by trade name,
trademark, manufacturer, or otherwise does not necessarily constitute or imply its
endorsement, recommendation, or favoring by the United States Government or any agency
thereof. The views and opinions of authors expressed herein do not necessarily state or reflect
those of the United States Government or any agency thereof.

\bibliographystyle{unsrt}
\bibliography{biblio}

\appendix

\section{Connection to Efficient Subsampled Natural Gradient Descent }\label{app:NGD-SMW} 
Recall that MinSR with Tikhonov regularization uses the update formula
\begin{equation}
d\theta = \bar{O}^T \paren{\lambda I + \bar{O} \bar{O}^T}^{-1} \bar{\epsilon}, \label{eq:minsr_damped}
\end{equation}
where $\lambda$ is the damping parameter. This is the same formula that arises from the ``simple linear algebra trick'' of Rende et al \cite{ren2019efficient}. We now introduce the efficient subsampled natural gradient method of Ren and Goldfarb \cite{ren2019efficient} and show how it is equivalent to MinSR in the VMC setting. The key idea of Ren and Goldfarb is that a damped Fisher matrix arising from a small minibatch can be viewed as a low-rank perturbation of the identity and is thus amenable to inversion using the Sherman-Morrison-Woodbury formula. Using the notation of the later work by Yang et al \cite{yang2022sketch}, the resulting update is
\begin{align*}
d\theta&= \frac{1}{\lambda} \paren{-I +  U \paren{\lambda I + U^T U}^{-1} U^T } g,
\end{align*}
where $g$ is the gradient estimator and the columns of $U$ contain the model gradients divided by the square-root of the number of samples used.

Now suppose, as is often the case, that the gradient of the loss function is simply a linear combination of the gradients at each input, with some input-dependent scalar weights. Then we can write the gradient estimator as $g = Uv$ for some column vector $v$ that depends on the sampled inputs. This allows us to simplify the NGD-SMW formula as follows:
\begin{align*}
d\theta &= \frac{1}{\lambda} \paren{-I + U \paren{\lambda I + U^T U}^{-1} U^T} U v\\
&=  \frac{1}{\lambda} \paren{ - U +  U \paren{\lambda I + U^T U}^{-1}U^T U } v\\
&= \frac{1}{\lambda} \paren{ - U  + U \paren{\lambda I + U^T U}^{-1} \paren{\lambda I +U^T U} - U \paren{\lambda I + U^T U}^{-1} \paren{\lambda I} }v \\
&=  - U \paren{\lambda I + U^T U}^{-1}v.
\end{align*}

Comparing with (\ref{eq:minsr_damped}), we see that this is equivalent to the MinSR formula when we identify $\bar{O}^T$ with $U$ and $\bar{\epsilon}$ with $v$.  Furthermore, this identification is the natural one, since the columns of $\bar{O}^T$ represent the gradients of the logarithm of the normalized wavefunction and the formula $g = \bar{O}^T \bar{\epsilon}$ holds in the VMC setting up to constant factors. Thus, the MinSR method can be viewed as a simplified way of implementing the efficient subsampled natural gradient method in the context of VMC. Furthermore, SPRING can be viewed as a potential improvement to subsampled natural gradient descent and may have applications outside the VMC setting. 

\section{Network Architecture and Hyperparameters \label{app:hyper}}
We list in Table \ref{tab:ferminet_arch} the hyperparameters of the FermiNet architecture that we use in all our experiments. We list in Table \ref{tab:vmc_hyp} the other settings used for our VMC training phase, in Table \ref{tab:inf} the settings used for our inference phase, and in Table \ref{tab:prelim_hyp} the settings used for our preliminary optimization. 

\begin{table}
    \centering
    \begin{tabular}{c || c}
    Hyperparameter & Value \\
    \hline \hline
    One-electron stream width & 256  \\
    Two-electron stream width &  16 \\
    Number of equivariant layers & 4 \\
    Backflow activation function & tanh \\
    Number of determinants & 16 \\
    Exponential envelope structure & Isotropic \\
    Determinant type & Dense\\
    \end{tabular}
    \caption{List of FermiNet architecture hyperparameters used for all experiments.}
    \label{tab:ferminet_arch}
\end{table}

\begin{table}
    \centering
    \begin{tabular}{c||c}
    Setting & Value \\
    \hline \hline
    Standard deviations for local energy clipping & 5 \\
     Number of walkers & 1000  \\
     MCMC burn-in steps & 5000 \\   
     Training iterations & 1e5 \\
     MCMC steps between updates & 10 \\
    \end{tabular}
    \caption{List of settings for the training phase.}
    \label{tab:vmc_hyp}
\end{table}

\begin{table}
    \centering
    \begin{tabular}{c||c}
    Setting & Value  \\    
    \hline \hline
    Number of walkers & 2000 \\
    MCMC burn-in steps for C,N,O & 5000 \\
    MCMC burn-in steps for N$_2$ & 1e4 \\
    Inference iterations & 2e4 \\
    MCMC steps between local energy measurements & 10 \\
    \end{tabular}
    \caption{List of settings for the inference phase.}
    \label{tab:inf}
\end{table}

\begin{table}
    \centering
    \begin{tabular}{c||c}
    Setting & Value  \\    
    \hline \hline
    Optimizer & KFAC \\
    Learning rate & 0.05 \\
    Kernel initializers & Orthogonal \\
    Bias initializers & Random normal \\
    Standard deviations for local energy clipping & 5 \\
    Number of walkers & 1000  \\
    MCMC burn-in steps & 5000 \\
    Training iterations for C,N,O & 1000 \\
    Training iterations for N$_2$ & 5000  \\
    MCMC steps between updates & 10 \\
    \end{tabular}
    \caption{List of settings for the preliminary optimization phase.}
    \label{tab:prelim_hyp}
\end{table}

\end{document}